\documentclass[aps,prb,superscriptaddress,showpacs,reprint]{revtex4-1}

\usepackage[utf8]{inputenc}
\usepackage{amsmath,amsfonts,amssymb}
\usepackage{pifont}
\usepackage{rotating}
\usepackage{enumerate}
\usepackage{graphicx}    
\usepackage{epstopdf}
\usepackage{color}
\usepackage{soul} 
\usepackage{subfigure}
\usepackage{ulem}

\usepackage{multirow,bigstrut}
\newsavebox{\measurebox}

\usepackage{hyperref}
\definecolor{dark-red}{rgb}{0.9,0.15,0.15}
\definecolor{dark-blue}{rgb}{0.15,0.15,0.4}
\definecolor{medium-blue}{rgb}{0,0,0.5}
\hypersetup{
    colorlinks, linkcolor={black},
    citecolor={dark-blue}, urlcolor={medium-blue}
}

\begin{document}

\title{Half-metallic ferromagnetism and Ru-induced localization in quaternary Heusler alloy CoRuMnSi}

\author{Y. Venkateswara}
\thanks{YV and DR contributed equally to this work.}
\affiliation{Magnetic Materials Laboratory, Department of Physics, Indian Institute of Technology Bombay, Mumbai 400076, India}

\author{Deepika Rani}
\thanks{YV and DR contributed equally to this work.}
\affiliation{Magnetic Materials Laboratory, Department of Physics, Indian Institute of Technology Bombay, Mumbai 400076, India}

\author{K. G. Suresh}
\affiliation{Magnetic Materials Laboratory, Department of Physics, Indian Institute of Technology Bombay, Mumbai 400076, India}

\author{Aftab Alam}
\email{aftab@phy.iitb.ac.in}
\affiliation{Department of Physics, Indian Institute of Technology Bombay, Mumbai 400076, India}

\begin{abstract}
	
We report a combined theoretical and experimental investigation of half-metallic ferromagnetism in equiatomic quaternary Heusler alloy CoRuMnSi. Room temperature XRD analysis reveals that the alloy crystallizes in L2$_1$ disorder instead of pristine Y-type structure due to 50\% swap disorder between the tetrahedral sites, i.e., Co and Ru atoms. Magnetization measurements reveal a net magnetization of 4 $\mu_B$ with Curie temperature, $\mathrm{T_C}$ of $\sim780$ K. Resistivity measurement reveals the presence of localization effect below 35 K while above 100 K, a linear dependence is observed.  Resistivity behavior indicates the absence of single magnon scattering, which indirectly supports the half-metallic nature. The majority spin band near the Fermi level clearly indicates the overlap of flat $e_g$ bands with sharply varying conduction bands that are responsible for the localization. In-depth analysis of the projected atomic d-orbital character of band structure reveals unusual bonding, giving rise to the flat $e_g$ bands purely arising out of Ru ions. Co-Ru swap disorder calculations indicate the robustness of half-metallic nature, even when the structure changes from Y-type to $L2_1$-type, with no major change in the net magnetization of the system. Thus, robust half-metallic nature, stable structure, and high $\mathrm{T_C}$ make this alloy quite a promising candidate to be used as a source of highly spin-polarized currents in spintronic applications.
\end{abstract}

\date{\today}

\maketitle

\section{Introduction}
Heusler alloys gained immense interest in the field of spintronics due to recent discoveries of several half-metals (HM), spin gapless semiconductors (SGS), magnetic semiconductors (MS) and spin semimetals (SSM). These materials produce spin-polarized current in their electrical conduction. These classes of materials also exhibit ferro-, ferri- or fully compensated ferri-/antiferro-magnetism. CoFeMnSi,\cite{PhysRevB.91.104408} CoFeCrGe,\cite{PhysRevB.92.224413} Co$_2$FeSi,\cite{PhysRevB.87.220402} etc. are ferromagnetic spintronic materials in which the spins of all the magnetic ions are aligned in the same direction and possess integer moments. Ferrimagnetic materials such as CoFeCrGa,\cite{PhysRevB.92.045201} Mn$_2$CoAl,\cite{PhysRevLett.110.100401} etc. contain two or more magnetic elements with their spins aligned in opposite directions, resulting in low integer moment but sufficiently exchange coupled to result in high Curie temperature ($T_C$). Fully compensated ferrimagnets (FCF) such as CrVTiAl,\cite{PhysRevB.97.054407} MnVCrAl,\cite{doi:10.1063/1.4998308} Mn$_{1.5}$FeV$_{0.5}$Al,\cite{doi:10.1063/1.5000351} and Mn$_3$Al\cite{PhysRevApplied.7.064036} are the special kind of materials from the family of ferrimagnets in which the net moment is close to zero, resembling antiferromagnets. Among the discovered spintronic materials, SGS gained a lot of interest due to their peculiar properties such as (i) spin-polarized electrons and holes, (ii) less or almost no energy required to excite carriers from valence band to conduction band unlike conventional semiconductors, (iii) possibility of Dirac or Weyl points at the Fermi level.\cite{doi:10.1063/1.5042604} Mn$_2$CoAl,\cite{PhysRevLett.110.100401} CoFeCrGa,\cite{PhysRevB.92.045201} CoFeMnSi,\cite{PhysRevB.91.104408} Ti$_2$MnAl,\cite{SINGH201815421} CrVTiAl\cite{PhysRevB.97.054407} are some of the predicted SGS materials from the Heusler family. One of the issue with SGS materials, however, is the sensitivity of its electronic structure. This is due to the gapless nature of one of the spin bands, which can easily change due to small perturbation such as (i) external pressure, (ii) disorder, etc. Most of the reported SGS materials have some disorder, which not only affects the SGS nature but also degrades the spin polarization. 

Even though there are a large number of Heusler alloys which are theoretically predicted to have high spin polarization and are experimentally verified, the measured spin polarization values are quite low. \cite{Picozzi_2007, doi:10.1063/1.4929252, PhysRevB.83.140409, PhysRevB.69.144413} The main cause of degradation of spin polarization is disorder, e.g., anti-site disorder, which destroys the half-metallic character. As such, in order to have a robust spintronic material, the primary objective would be to minimize the disorder. Unfortunately, in the quaternary Heusler family, only a few alloys have been reported to crystallize in perfectly ordered Y-type structure. A key question, therefore, is how to achieve a minimal disorder.

It is also well-known \cite{Graf-simplerules-Heusler-PSSC-review, PhysRevB.92.224413} that, to minimize the disorder, the element of the octahedral site should be less electronegative than that at the tetrahedral site. A large number of Heusler alloys which show a considerable amount of disorder, e.g., CoFeMnGe,\cite{PhysRevB.96.184404} CoFeMnSi, \cite{PhysRevB.91.104408} MnCrVAl,\cite{doi:10.1063/1.4972797} CrVTiAl \cite{PhysRevB.97.054407} etc., have more than one element with nearly equal (small) electronegativity, and hence compete to occupy the octahedral site. Thus, if the elements with similar electronegativity values are the constituents of a Heusler alloy, it results in disorder.\cite{PhysRevB.92.224413, PhysRevB.97.054407, doi:10.1063/1.4943306} One way to minimize the disorder is to replace one such 3d- element by a 4d or 5d- element. This can be observed in CoRhMnGe,\cite{PhysRevB.96.184404} CoRhMnSn,\cite{Alijani_2012} CoRuFeX (X=Si, Ge),\cite{BAINSLA2015631} Rh$_2$MnGe,\cite{Klaer_2009} Ru$_2$MnGe \cite{KANOMATA20061} etc. All these alloys exhibit L2$_1$ structure without any disorder between octahedral atoms. It should be noted that the electronegativity values for Fe and Mn are 1.83 and 1.55 respectively, whereas Ru and Rh have higher electronegativity of 2.20. \cite{Graf-simplerules-Heusler-PSSC-review} CoFeMnSi is predicted to be SGS by us in the past, with DO$_3$ disorder. \cite{PhysRevB.91.104408} This disorder is due to Fe, which competes with Mn and occupies both octa and tetrahedral sites, as revealed by Mossbauer spectroscopy. One of the main motivation of this paper is to replace Fe by Ru (4d- element), which is expected to improve the crystal structure and magnetic properties.

\begin{figure}[t!]
\centering
\includegraphics[width=\linewidth]{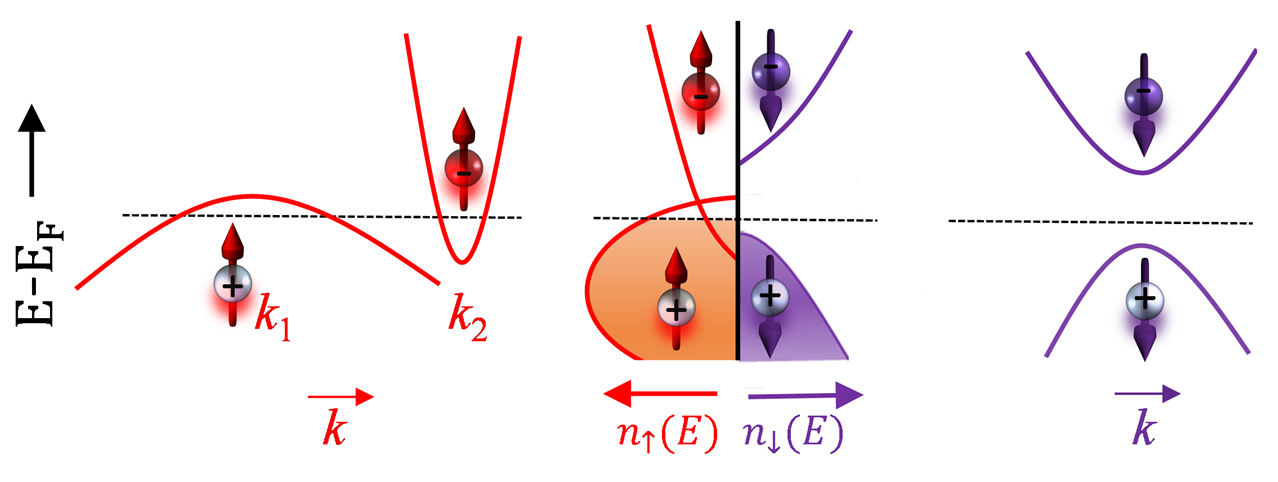}
\caption{Spin resolved band structure for a localized half metal.(left) spin up band with a small overlap of flat valence and sharp conduction bands at $\mathbf{k}_1$ and $\mathbf{k}_2$ respectively. (Middle) spin resolved partial density of states for valence and conduction bands, (right) spin down bands having finite gap.}
\label{fig:schem-WSM}
\end{figure}

Ideally one of the spin band is gapless while the other is gapped in SGS. The gapless nature can either arise from (i) flat valence and sharp conduction bands or (ii) flat valence and conduction bands, or (iii) sharp valence and flat conduction bands, or (iv) sharp valence and conduction bands. However, if the spin band deviates from gapless nature resulting in small overlap of valence and conduction bands, its properties greatly depend on the type of bands involved, as discussed above. The scenarios (ii) and (iv) remain close to SGS up to some extent before becoming SSM, whereas the scenarios (i) and (iii) give localized half-metals. Figure \ref{fig:schem-WSM} shows a schematic of the spin-resolved band structure for the scenario (i), which is a localized half-metal (LHM). Notice from the band dispersion that the valence band contains heavier holes while conduction band contains lighter electrons in the spin-up band, whose partial density of states is shown in Fig. \ref{fig:schem-WSM}(middle). The spin-up band resembles that of conventional localized metals. For LHM, temperature coefficient of resistivity ($\partial\rho/\partial T$) is positive, except at low temperatures at which $\partial \rho/\partial T$ is negative as both heavy holes and light electrons contribute to the electrical conductivity due to less dominant phonons. At higher temperatures, light carriers can excite to higher energies and dominate in electrical conductivity over heavy carriers due to thermal phonons. Heavy carriers get trapped and recombine with some lighter electrons, and so their contribution is negligible in the variation of electrical resistivity. To a certain extent, CoRuMnSi belongs to this scenario, as explained in detail in section IV. 

Theoretical simulations on CoRuMnSi have been partially performed in previous reports by Kundu \textit{et. al.}\cite{Kundu-CoRuMnSi-theory-Scientificreports} and Kalaf \textit{et. al.}\cite{Khalaf-CoRuMnSi-theory-RSC} However, neither of these dealt with the localized half-metallic nature. Also, no experimental reports on CoRuMnSi alloy exist in the literature so far. 

\section{Experimental and theoretical details}

\subsection{Experimental details}

Polycrystalline CoRuMnSi (CRMS) alloy was synthesized by arc-melting the stoichiometric weights of constituent elements Co, Ru, Mn, Si (with purity greater than 99.99\%) in high purity argon atmosphere. To compensate for the weight loss due to Mn evaporation during melting, 2\% extra Mn was taken. \cite{PhysRevB.96.184404} The ingot was melted several times, and the final weight loss was less than 1\%. The alloy was annealed at 800 $^oC$ for 7 days followed by furnace cooling. Room-temperature X-ray diffraction patterns were taken using Cu-K$_\alpha$ radiation with the help of Panalytical X-pert diffractometer. Phase purity and crystal structure analysis were done using FullProf suite.\cite{RR, RODRIGUEZCARVAJAL199355} Magnetization isotherms at 3 K and 300 K were obtained using a vibrating sample magnetometer (VSM) attached to the physical property measurement system (PPMS) (Quantum Design) for fields up to 50 kOe. High-temperature magnetization data were collected in the field of 100 Oe using oven option in VSM. Electrical resistivity measurements were done using the four-probe method in PPMS with an excitation current of 5 mA.

\subsection{Theoretical details}

Ab initio simulations are performed by using a spin-resolved density functional theory (DFT) implemented within Vienna Ab initio Simulation Package (VASP) \cite{PhysRevB.54.11169} with a projected augmented-wave basis. \cite{PhysRevB.59.1758} The electronic exchange-correlation potential due to Perdew, Burke, and Ernzerhof (PBE) is used within the generalized gradient approximation (GGA) scheme. A $24^3$ k mesh is used to perform the Brillouin zone integration within the tetrahedron method. A plane wave energy cutoff of 500 eV is used for all the calculations. All the structures are fully relaxed (cell volume, shape, and atomic positions of constituent atoms), with total energies (forces) converged to values less than $10^{-6}$ eV (0.01 eV/\AA).

The quaternary Heusler alloys XX$'$YZ\cite{Graf-simplerules-Heusler-PSSC-review} crystallizes in Y type structure with prototype LiMgPdSn (space group \# 216), whose primitive cell contains four atoms at the Wyckoff positions, 4a(0, 0, 0), 4b(1/2, 1/2, 1/2), 4c(1/4, 1/4, 1/4), and 4d(3/4, 3/4, 3/4). The three possible nondegenerate crystallographic configurations using 4 atom primitive cells are (by keeping Z at 4a(0, 0, 0) site)
\begin{enumerate}[I.]
\item   X at 4d, X$'$ at 4c, and Y at 4b sites,
\item	X at 4b, X$'$ at 4d,  and Y at 4c sites, 
\item	X at 4d, X$'$ at 4b, and Y at 4c sites
\end{enumerate}
respectively.

We have also simulated the effect of L2$_1$ disorder that occurs due to the 50\% disorder of tetrahedral site atoms. A $2\times2\times2$ supercell involving 32-atoms is formed from a four-atom primitive cell of the energetically most stable configuration. The Brillouin zone sampling used for this calculation is $6^3$ $k-$mesh. Xcrysden\cite{kokalji} and Vesta\cite{Mommadb5098} programs were used in visualizing crystal structures, Fermi surfaces.

\section{Experimental results}

\subsection{Crystal structure}

\begin{figure}[tb!]
\centering
\includegraphics[width=0.8\linewidth]{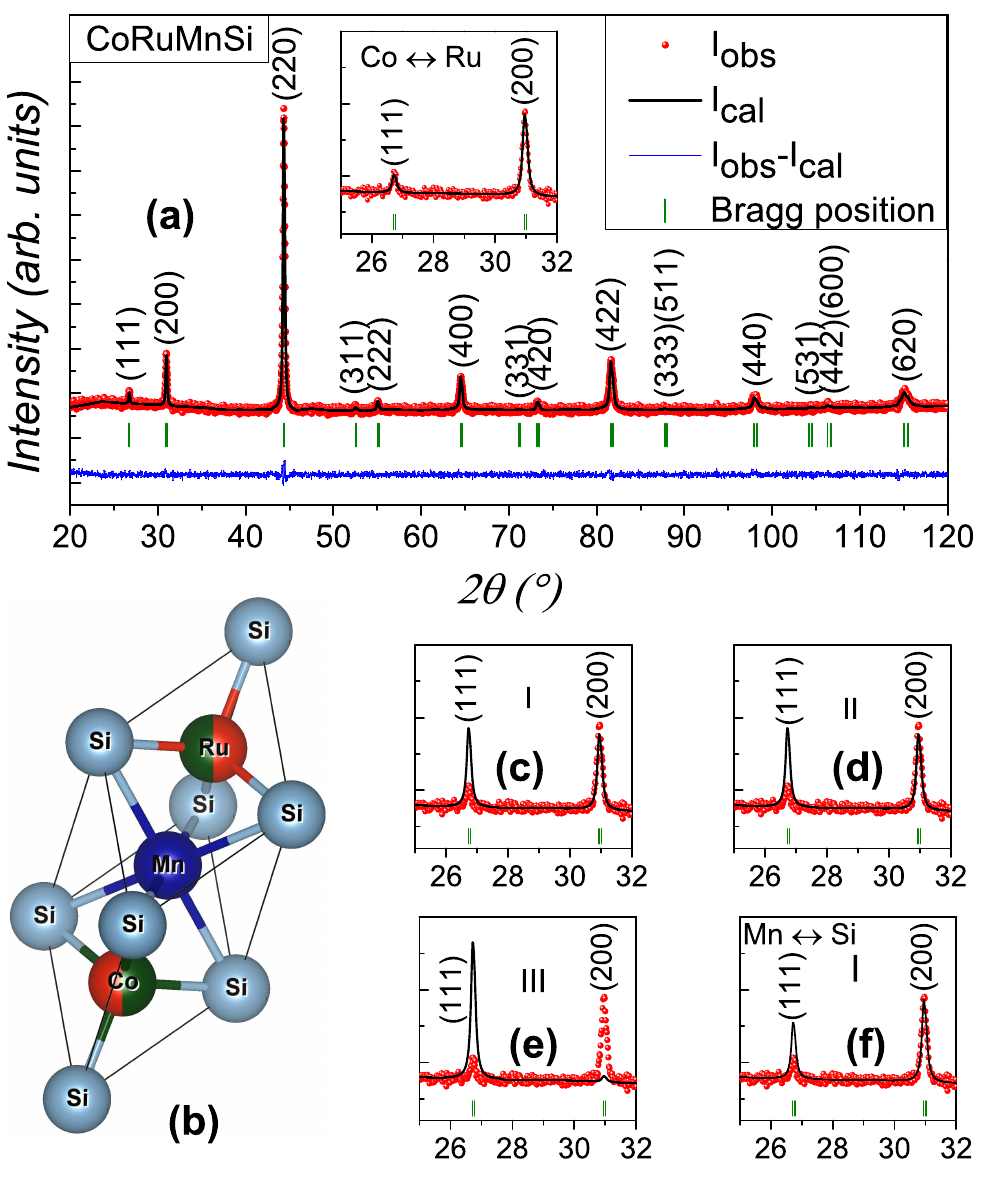}
\caption{Rietveld refined room temperature XRD pattern of polycrystalline CoRuMnSi. (a) XRD pattern with 50 \% disorder at tetrahedral site atoms Co and Ru. Inset shows a zoomed view near (111) ad  (200) superlattice reflection peak. (b) Primitive unit cell corresponding to the fitted structure in (a). Rietveld refinement corresponding to configurations I, II and III are shown in (c), (d) and (e) respectively. (f) Refinement considering 50 \% disorder between octahedral site atoms Mn and Si in configuration I.}
\label{fig:XRD-CoRuMnSi}
\end{figure}

Room temperature powder XRD pattern of CoRuMnSi shown in Fig. \ref{fig:XRD-CoRuMnSi}(a) can be indexed in LiMgPdSn type structure. The calculated lattice parameter is 5.78 \AA. However, due to 50\% disorder between tetrahedral site atoms (Co and Ru), the crystal symmetry changes to L2$_1$ type. Among the superlattice peaks, intensities of (111) and (200) play a great role in understanding the disorder.\cite{PhysRevB.99.104429} Rietveld refinement considering any of the pure configurations (I, II, II) did not fit well (see Fig. \ref{fig:XRD-CoRuMnSi}(c)-(e)). Intensities of odd reflections did not match for the configurations I and II while for configuration III, intensities for both the odd and even super-lattice reflections did not match well. This suggests that either configuration I or II will be the preferred crystal structure with disorder between octahedral site atoms or tetrahedral site atoms. Refinement considering 50\% disorder of octahedral site atoms, Mn and Si, in the configuration I did not fit for the odd reflections (see Fig.\ref{fig:XRD-CoRuMnSi}(f)). The other possibility, i.e., having 50\% disorder of tetrahedral site atoms, Co and Ru, in the configuration I fits the best, and the refinement is shown in Fig. \ref{fig:XRD-CoRuMnSi}. Inset of Fig. \ref{fig:XRD-CoRuMnSi}(a) shows the zoomed-in view near the super-lattice reflections (111) and (200). Its primitive crystal cell is shown in Fig. \ref{fig:XRD-CoRuMnSi}(b). Refinement considering B2 disorder, i.e., a simultaneous disorder of the octahedral site atoms (Mn and Si) and the tetrahedral site atoms (Co and Ru) completely suppresses the odd reflections in any of the configurations. The difference of atomic form factors of Co and Mn is very negligible in comparison to that of Ru and Ge, and hence any disorder refinement does not distinguish configurations I and II. However, based on the empirical relation between electronegativity and site occupancy, Co atoms will not occupy the octahedral site. Therefore, even though the refinement considering 50\% disorder of tetrahedral site atoms Mn and Ru in configuration II fits well, this configuration should be discarded. Hence, it can be assumed that CoRuMnSi crystallizes in L2$_1$ disordered structure due to 50\% disorder of tetrahedral site atoms Co and Ru in the configuration I. Compared to CoFeMnSi, CoRuMnSi (CRMS) only has disorder between tetrahedral sites which results in $L2_1$-type structure. Disorder between tetrahedral sites is common in quaternary Heusler alloys CoRhMnZ, \cite{Alijani_2012} as their bond lengths are large and their surrounding chemical environment is identical.

\subsection{Magnetization}

Magnetic moment of quaternary Heusler alloys can be calculated using the following relation, which is based on the Slater Pauling rule,\cite{PhysRevB.83.184428,Slat1,Paul1} 
\begin{equation}
M_S = N_v - 24, \mathrm{in \ \mu_B/f.u.}
\end{equation}
where $M_S$ is the saturation magnetization and $N_v$ is the total number of valence electrons in the alloy.

\begin{figure}[b!]
\centering
\includegraphics[width=\linewidth]{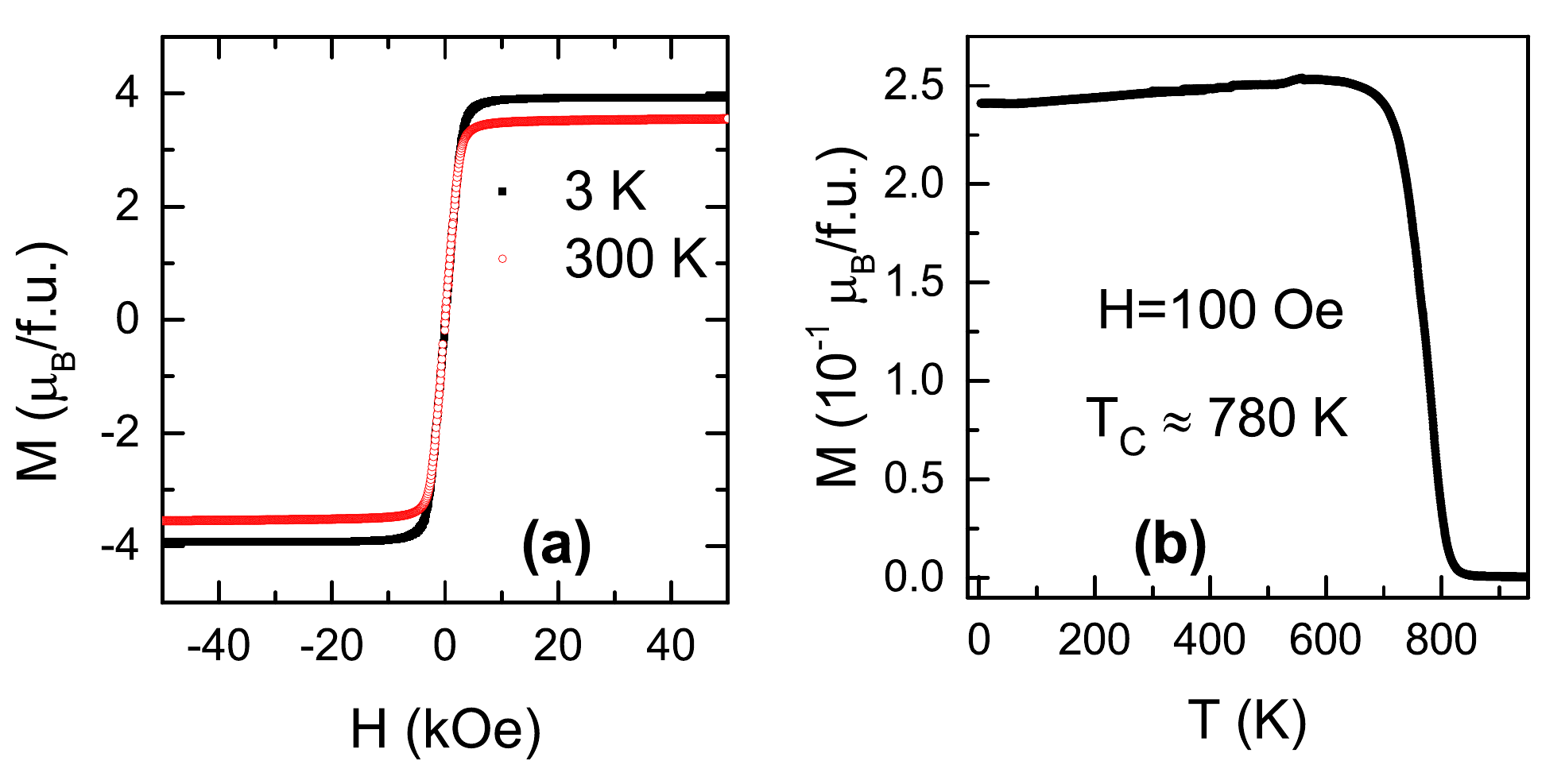}
\caption{(a) Isothermal magnetization curves at 3 K and 300 K for CoRuMnSi. (b) Magnetization vs. temperature data in field warming mode in 100 Oe.}
\label{fig:MT-MH}
\end{figure}

\begin{table*}
	\centering
	\caption{Fitting parameters for electrical resistivity vs. $T$ in zero field for $\mathrm{CoRuMnSi}$ alloy. $\rho_{01}$ = 35.85 $\mu\Omega$ cm and $\rho_{02}$ = 35.62 $\mu\Omega$ cm.}
	
	\begin{tabular}{c c c c}
		\hline \hline & & \\
		Temperure Region & Fitting Equation & \multicolumn{2}{c}{{Fitting Parameters}}\\
		\hline & & \\
		4 K $<$ T $<$ 35 K & $\rho(T) = \rho_{01}-AT^{1/2}$ &$ A = [0.030(1)]$$\mu\Omega \mathrm{\:cm\:K^{-1/2}}$ &\\
		\hline& & \\
		 35 K $<$ T $<$ 100 K & $\rho(T) = \rho_{02} + BT^n $ & $B= [4.1(8)]\times{10^{-6}}$ $\mu\Omega \mathrm{\:cm\:K^{-n}}$ & $ n=2.59(4)$ \\
		\hline& & \\
		 100 K $<$ T $<$ 300 K & $\rho(T) = \rho_{02} + BT^n $ & $B= [1.02]\times{10^{-2}}$ $\mu\Omega \mathrm{\:cm\:K^{-n}}$ & $ n=1.066(4)$ \\
		\hline \hline
	\end{tabular} 
	\label{tab5} 
\end{table*}

Figure \ref{fig:MT-MH}(a) shows the variation of magnetization with the field at 3 K and 300 K of CRMS. It has saturation value nearly 3.92 $\mathrm{\mu_B/f.u.}$ at 3 K, which is close to 4.00 $\mathrm{\mu_B/f.u.}$ as predicted by Slater Pauling rule. It falls to 3.6 $\mathrm{\mu_B/f.u.}$ at 300 K. It is found to be a soft ferromagnetic with negligible hysteresis. Figure \ref{fig:MT-MH}(b) shows the temperature variation of magnetization in a field of 100 Oe. The Curie temperature is found to be nearly 780 K, which is much higher than that CoFeMnSi (620 K). The higher magnetic ordering temperature could be due to less disorder observed in the alloy, which in turn affects the exchange interactions.

\subsection{Resistivity Measurements}

\begin{figure}[t!]
	\centering
	\includegraphics[width=0.6\linewidth]{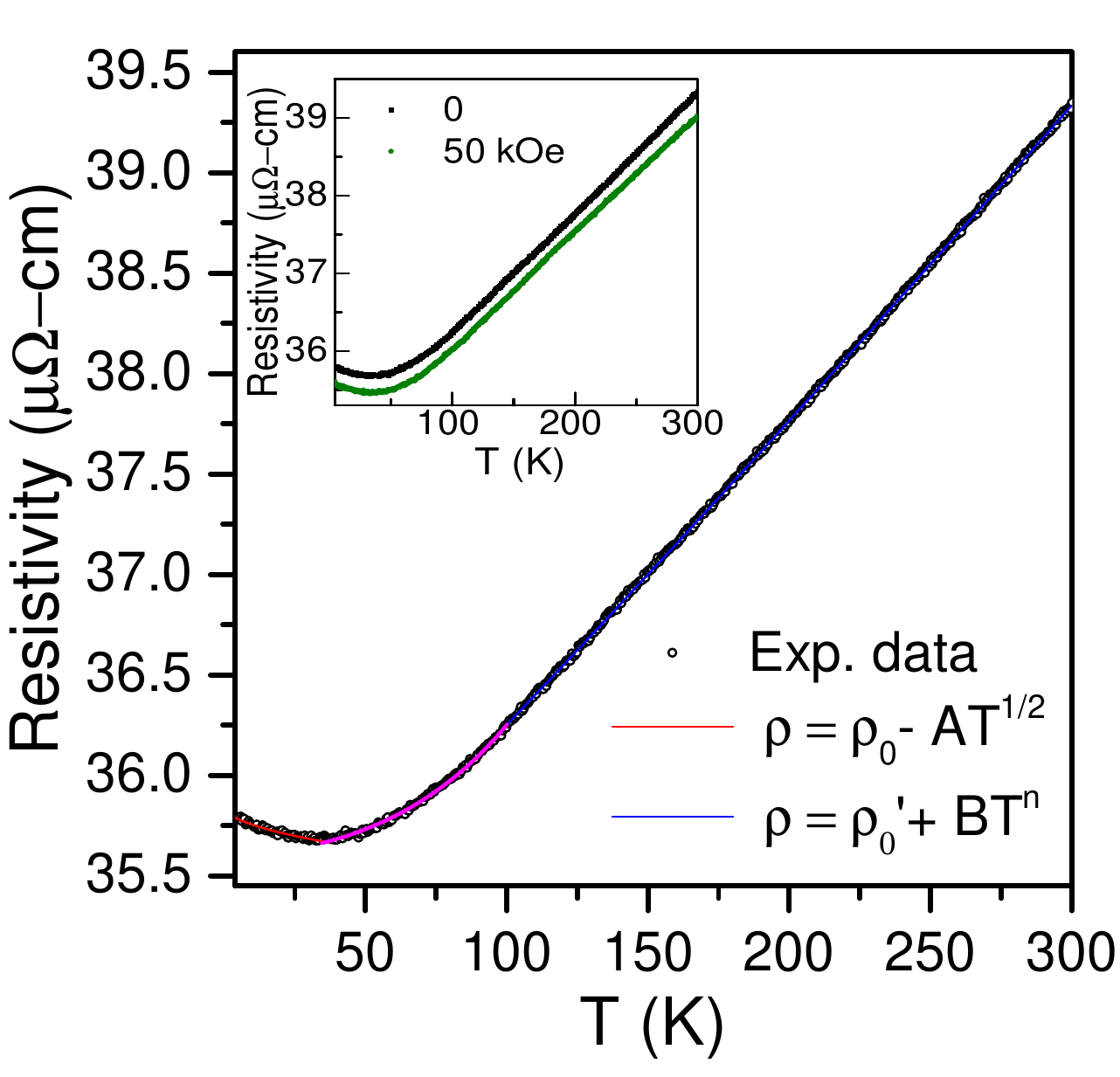}
	\caption{Temperature dependence of electrical resistivity for CoRuMnSi in zero field along with the fit to Eq. 2 (red) and Eq. 3 (blue). The inset shows the resistivity variation with temperature in 0 and 50 kOe fields.}
	\label{fig:CoRuMnsi_res_fit}
\end{figure}

Variation of electrical resistivity $(\rho)$ with temperature can be used to get an insight into the band structure and the carrier excitations. For example, absence of T$^2$ term (due to single-magnon scattering) in many Heusler alloys have been considered as a signature of half-metallic nature.\cite{PhysRevApplied.10.054022, PhysRevB.96.184404, Aftab_2012, doi:10.1063/1.4813519} This is because, in half-metals, the spin-flip scattering is not possible due to the unavailability of minority spin carriers at the Fermi level $(E_F)$.\cite{1989JPCM1.2351O,Kubo-1972}

Figure \ref{fig:CoRuMnsi_res_fit} shows the measured and fitted temperature dependence of resistivity in zero field. The inset shows the variation of resistivity with the temperature in two different fields (0 and 50 kOe), and one can notice that the sample exhibits a small negative magneto-resistance. An upturn in resistivity data below 35 K (see Fig. \ref{fig:CoRuMnsi_res_fit}), hints at the possibility of localization. A similar behavior is reported in other Heusler alloys. \cite{doi:10.1063/1.4862966, doi:10.1063/1.4902831,PhysRevApplied.10.054022} The localization phenomenon results in $\mathrm{T^{1/2}}$ term in resistivity. Thus, to have a better understanding, we carried out the resistivity data analysis in different temperature regimes considering different dominating scattering mechanisms.
 For 4 K $<$ T $<$ 35 K i.e., below the resistivity minimum, the data fits well to the equation
 \begin{equation}
 \rho(T) = \rho_0+ \rho(T) = \rho_{0} -AT^{1/2}
 \end{equation}
 In the temperature range, 35 K $<$ T $<$ 300 K, however, the resistivity data is found to fit with the power law,
 \begin{equation}
 \rho(T) = \rho_{0}+ \rho(T) = \rho_{0} + BT^n
 \end{equation}

The values of the best-fitted parameters are listed in Table \ref{tab5}. The value of n is found to be 2.59 in the temperature range 35 K $<$ T $<$ 100 K. A similar value of n was also observed for NiMnSb alloy. \cite{doi:10.1063/1.1739293} For 100 K $<$ T $<$ 300 K resistivity varies almost linearly with temperature with $n=1.07$ which is attributed to the electron-phonon scattering. The $T^{1/2}$ dependence at low temperatures (T $<$ 35 K) clearly indicates the presence of localization, which is also evident from the spin-resolved band structure (see Sec.IV for more details). 
Thus, the absence of $T^2$ term in the resistivity data for CoRuMnSi alloy indirectly supports the half-metallic nature.

\begin{figure*}[ht!]
\centering
\includegraphics[width=0.9\linewidth]{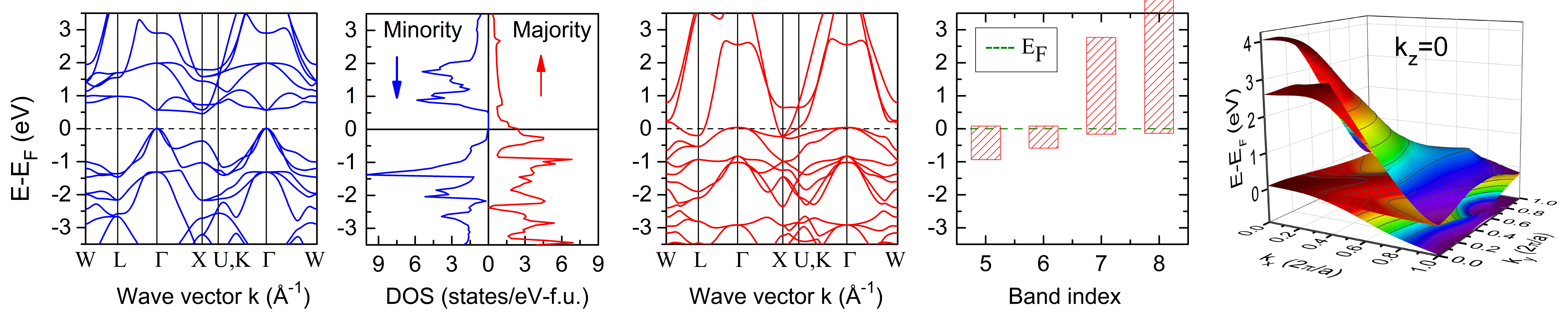}
\caption{Spin resolved density of states and band structure for CoRuMnSi at its equilibrium lattice parameter. The dispersion and band widths of four spin up bands that are crossing the $E_F$ are shown in two right most figures respectively.}
\label{fig:DB-CoRuMnSi-GGA}
\end{figure*}

\section{Theoretical Results}
Various magnetic states were examined by considering different initial spin (nonmagnetic, ferro and ferrimagnetic) arrangements for all the three configurations, as mentioned earlier. The results of the structural optimization are displayed in Table \ref{tab:magbeh-diff-configs-CoRuMnSi}. One can notice that the configuration I has the lowest energy among all the configurations with the ferromagnetic alignment of ions and hence is the ground state. The nonmagnetic(NM) state was found to be much higher in energy ($E_{NM}-E_{FM}$ = 0.3 eV/atom) as compared to the ferromagnetic(FM) state for the configuration I. Interestingly, the energy difference among the three configurations are sufficiently large, which indirectly indicates less possibility for disorders in this alloy. The ground state configuration I can be understood on the basis of empirical relation involving electronegativity, which suggests that the least electronegative ion occupies the octahedral site.\cite{Graf-simplerules-Heusler-PSSC-review, Venkateswara-FeRhCrGe-arxiv1902.01593v2} In CoRuMnSi, Mn has the least electronegativity and hence favors the octahedral site while the intermediate electronegative Co and Ru prefer tetrahedral sites. Dominant magnetization arise from Mn atom (2.96 $\mathrm{\mu_B/f.u.}$) whereas, Co and Ru contribute nearly 0.92 $\mathrm{\mu_B/f.u.}$ and 0.13 $\mathrm{\mu_B/f.u.}$ respectively. The relaxed lattice parameter, $a_{0}$ (5.79 \AA) is quite close to the experimental value, $a_{exp}$ (5.78 \AA). The calculated moment is 4.00 $\mu_B$, and it follows the SP rule and is also consistent with the experimentally observed value.
\begin{figure}[ht!]
	\centering
	\subfigure[]{\includegraphics[width=0.5\linewidth]{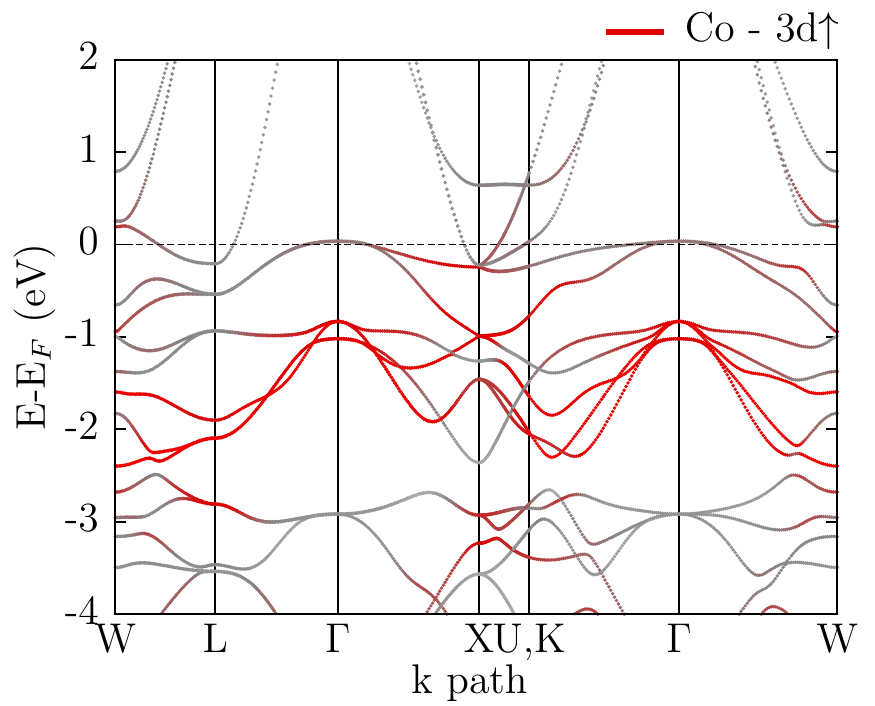}}%
	\subfigure[]{\includegraphics[width=0.5\linewidth]{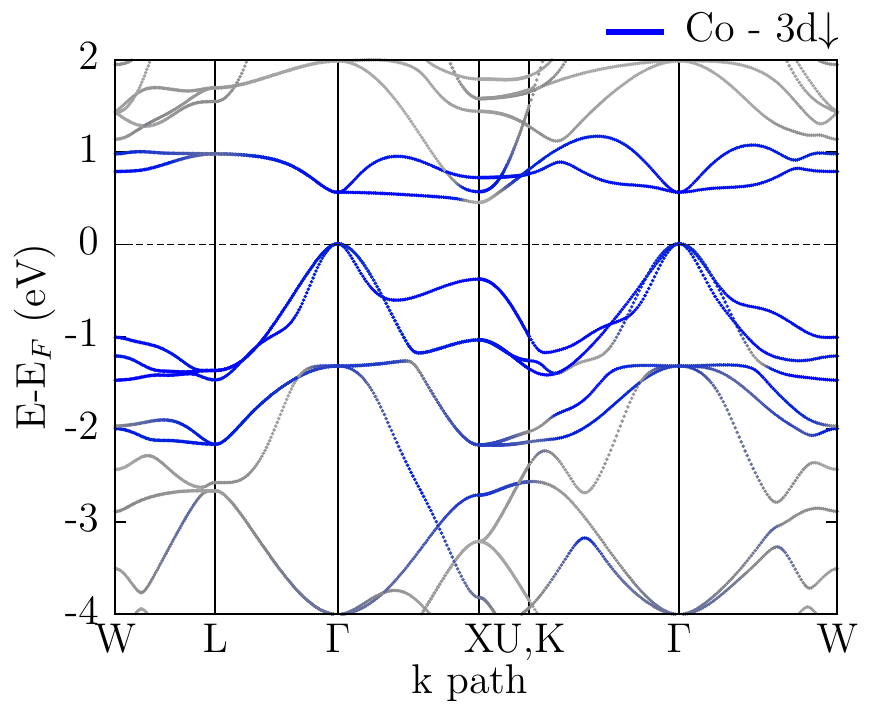}}
	
	\subfigure[]{\includegraphics[width=0.5\linewidth]{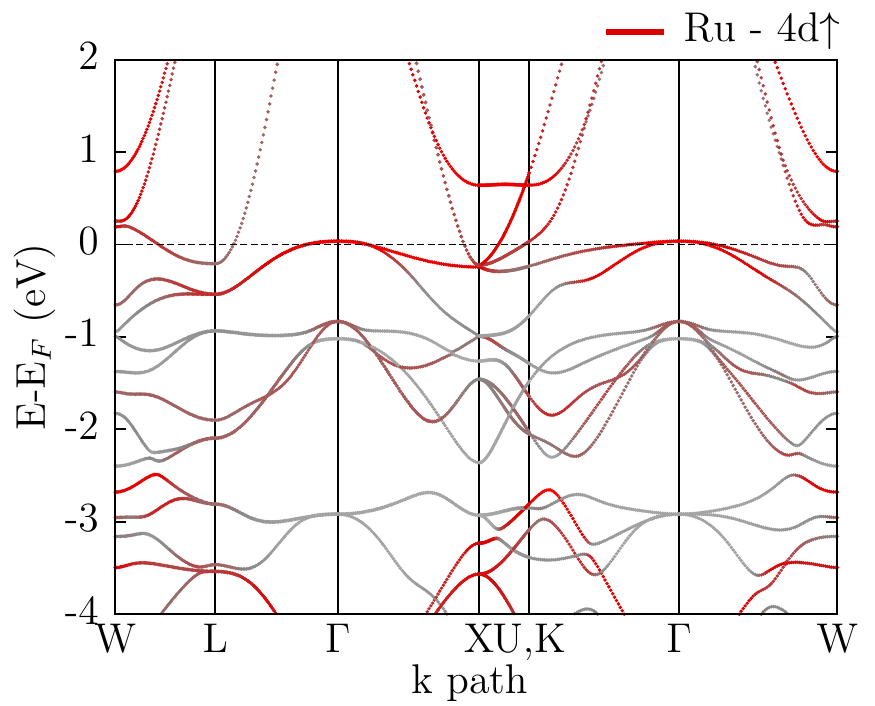}}%
	\subfigure[]{\includegraphics[width=0.5\linewidth]{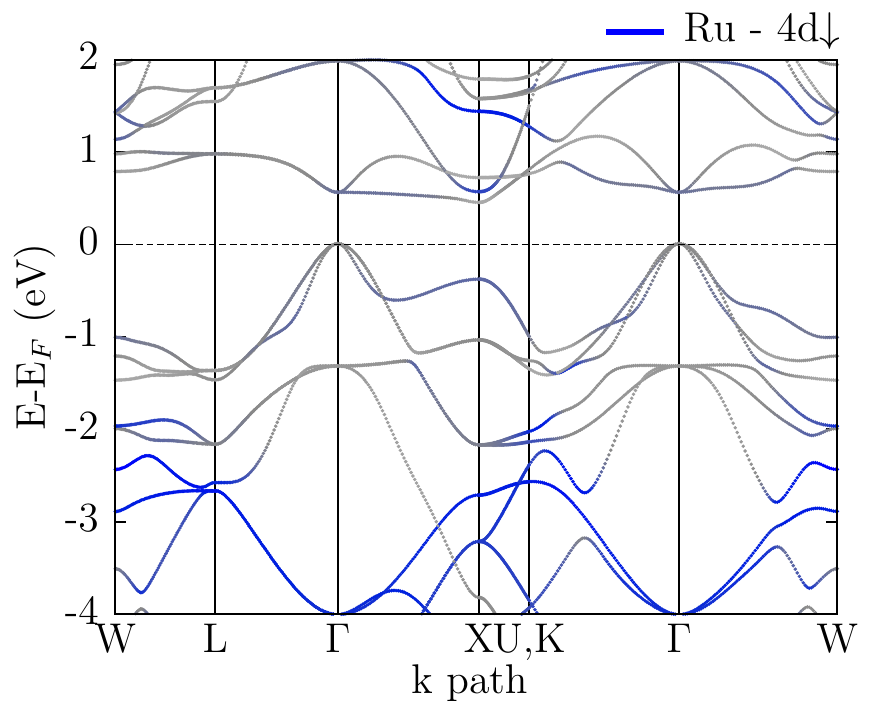}}
	
	\subfigure[]{\includegraphics[width=0.5\linewidth]{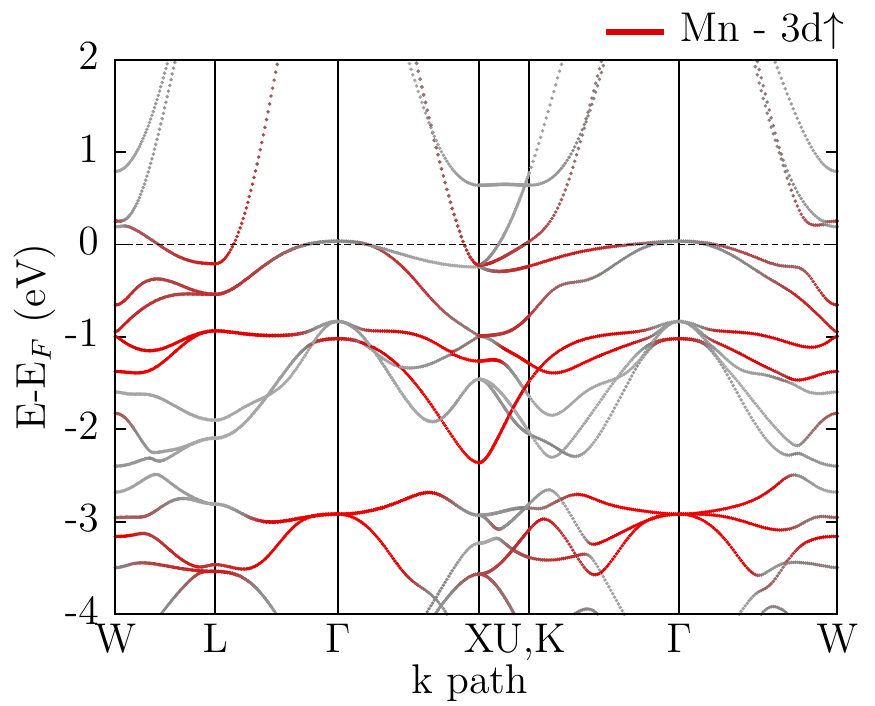}}%
	\subfigure[]{\includegraphics[width=0.5\linewidth]{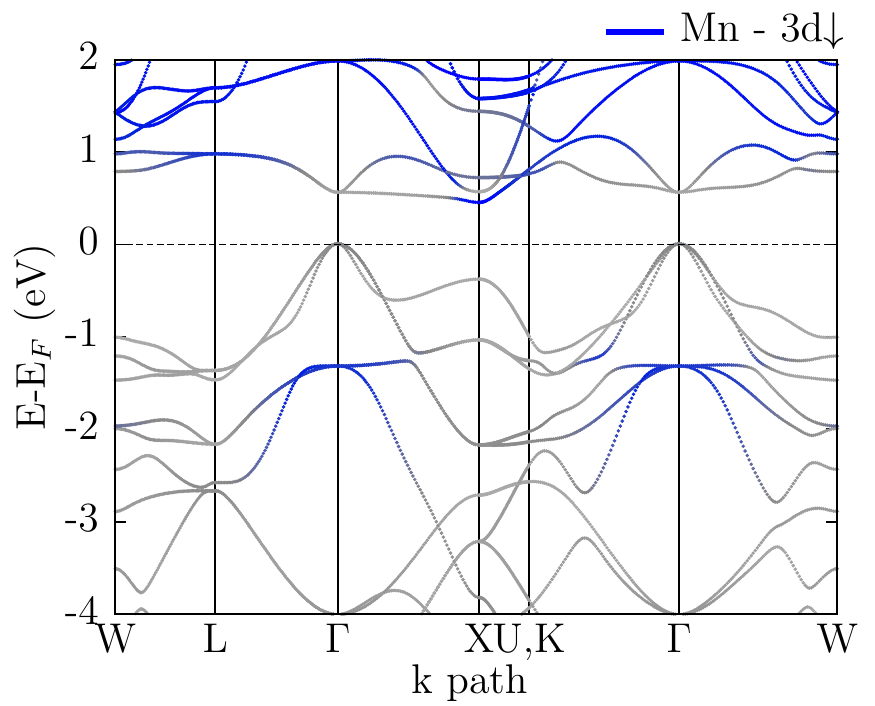}}
	
	\caption{For CoRuMnSi, band structure with projected d-orbital character for (a) Co, spin-up, (b) Co, spin-down (c) Ru, spin-up, (d) Ru, spin-down, (e) Mn, spin-up, and fb) Mn, spin-down.}
	\label{fig:proj-orb-CoRuMnSi}
	
\end{figure}
\begin{figure}[htp!]
	\centering
	\includegraphics[width=0.6\linewidth]{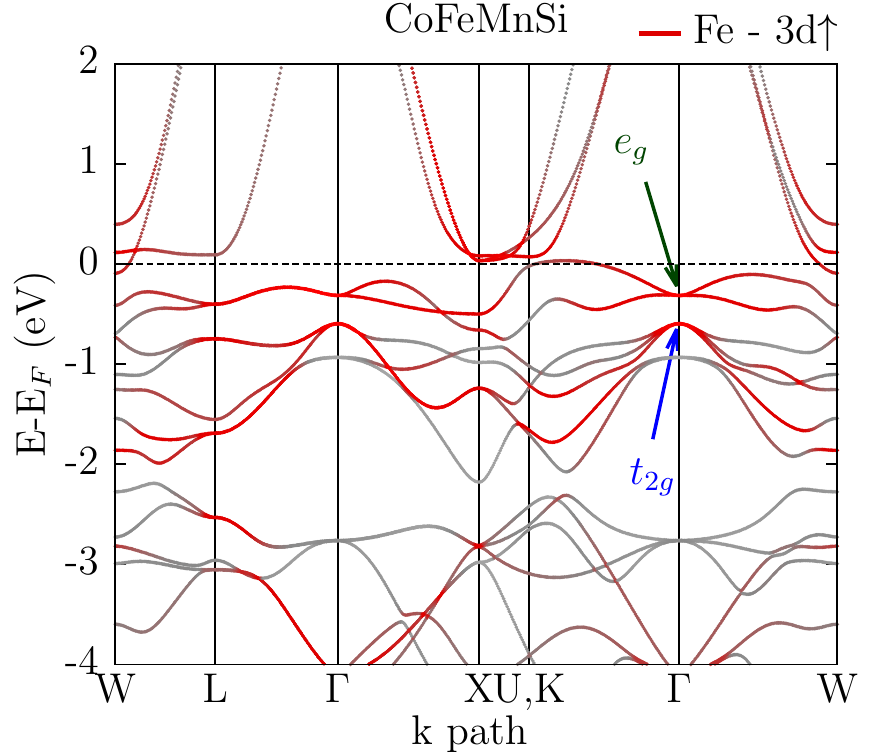} 
	\caption{d-orbital character of Fe ions in spin up band structure of CoFeMnSi for understanding the effect of Fe ions resulting in SGS nature.}
	\label{fig:proj-3dFe-CoFeMnSi}
\end{figure}
\begin{table}[h!]
\centering

\caption{Relaxed lattice parameter ($a_0$), atom-projected magnetic moments, total moments ($\mu_\mathrm{B}$) and relative energy ($\Delta E_{rel}$) with respect to the type I configuration for the three configurations I, II and III of CRMS.}

\begin{tabular}{l c c c c c c}
\hline \hline
& & & & & & \\
Type& $\ $ $a_0$ (\AA) $ \ $  &  $m^{\mathrm{Co}}$ & $\ $ $m^{\mathrm{Ru}}$ $\ $  &  $m^{\mathrm{Mn}}$  & $\ $ $m^{\mathrm{Total}}$ $ \ $ & $\Delta E_{rel}$(eV/atom) \\ & & & & & &\\ \hline \\
I    & 5.79   & 0.92 	&     0.13  	&	2.96 	& 4.00 		& 0.00   \\ & & & & & & \\
II   & 5.72   & 0.57 	& 	  0.10   	&  -0.29	& 0.39		& 0.23  \\ & & & & & & \\
III  & 5.75   & 1.26	& 	  0.79 		& 	0.74 	& 2.79 		& 0.34   \\ & & & & & & \\
\hline \hline
\end{tabular}

\label{tab:magbeh-diff-configs-CoRuMnSi}
\end{table}
\begin{figure}[h!]
	\centering
	\includegraphics[width=\linewidth]{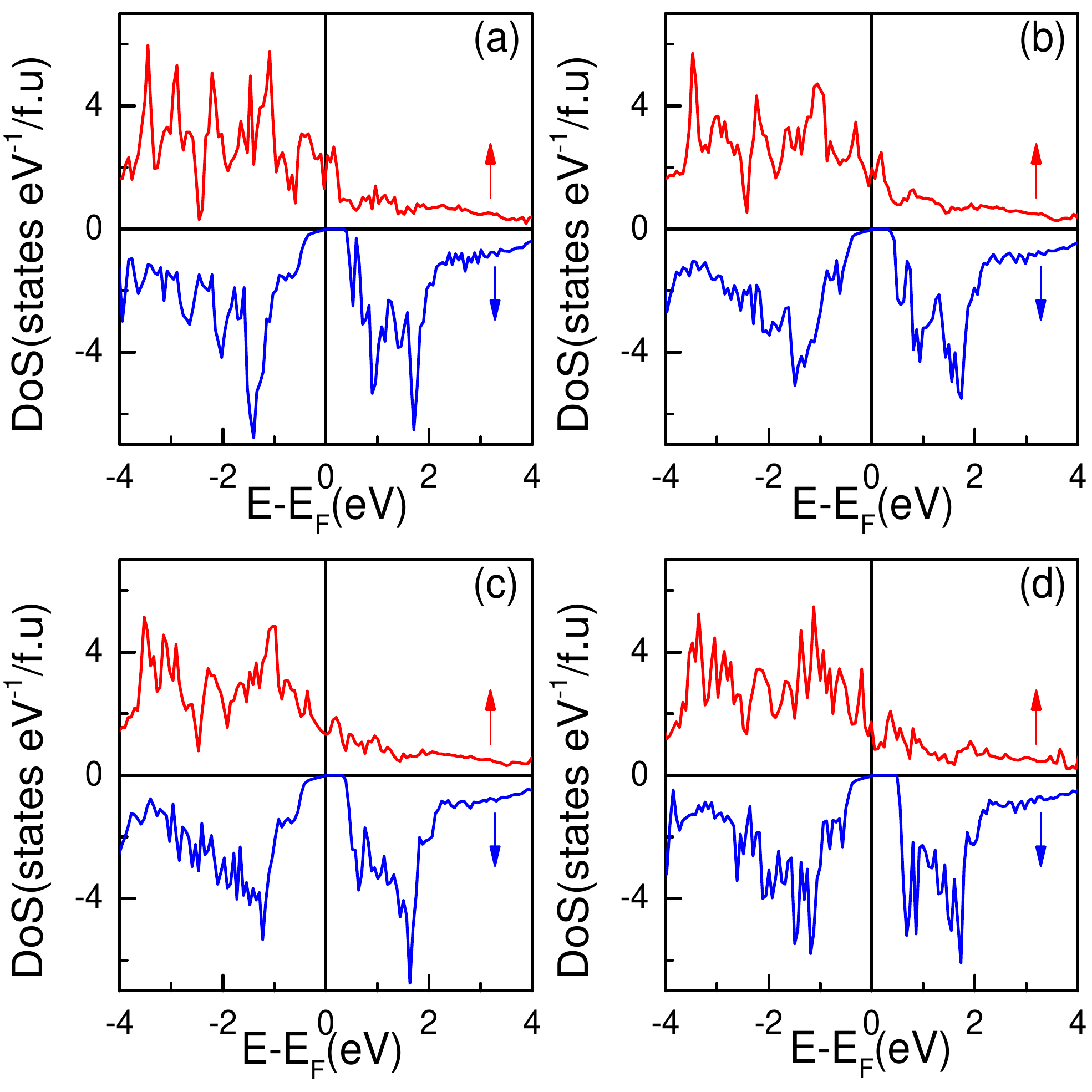}
	\caption{Spin-resolved DoS for CoRuMnSi with (a) 12.5\%, (b) 25 \%,(c) 37.5 \% and (d) 50 \% Co-Ru swap disorder.}
	\label{dis}
\end{figure}

Figure \ref{fig:DB-CoRuMnSi-GGA} shows the spin-resolved density of states, band structures, band widths for the spin up band, and 3D band dispersion for configuration I of CoRuMnSi at its relaxed lattice parameter $a_0$. The alloy manifests electronic structure of a fully spin-polarized half-metal. It has a finite gap of nearly 0.50 eV in the minority spin band. The two right most figures represent the band widths of a few spin up bands near the Fermi level nad 3D band dispersion respectively. One can notice that bands 5 and 6 have localized $e_g$ character while bands 7 and 8 are delocalized in nature. The localized bands are flat, causing heavy hole pockets around $\Gamma$ point (notice the surface dispersion plot) while delocalized bands have large band width with sharply varying nature, and are causing electron pockets around L and X points. Hence, as discussed in the introduction, the carriers from the bands 5 and 6 are less mobile compared to bands 7 and 8. Therefore, the overlap of flat valence bands 5 and 6 with sharply varying conduction bands 7 and 8 resembles the localized half-metallic nature and is consistent with the observed resistivity measurements. Hence, we conclude that the observed localization effect in CoRuMnSi is an intrinsic property of the alloy arising from its electronic band structure.\\
\begin{table*}
	\caption{Magnetic moments ($\mu_\mathrm{B}$) of perfect and disordered CoRuMnSi supercell ($2\times2\times2$). Here Co1,2,3,4,5,6,7,8 represents the eight Co atoms in a $2\times2\times2$ supercell and $\mathrm{X_d}$ refers to defect atoms. $\mathrm{Co_{Ru}}$ represents the moment of Co sitting at Ru site and $\mathrm{Ru_{Co}}$ represents the moment of Ru atom sitting at Co site.}
	\begin{tabular}{c| c |c| c| c |c |c}
		\hline \hline
		& & & &  & \\
		System & Co1,2,3,4,5,6,7,8 & Ru1,2,3,4,5,6,7,8 & Mn1,2,3,4,5,6,7,8 & $\mathrm{X_d}$ & $\mathrm{m_{total}}$  & $\mathrm{m_{total}}$   \\
		& & & & &($\mu_B$/cell) & ($\mu_B/f.u.$)\\
		\hline\\
		Perfect & 0.92 & 0.12 & 2.96 & & 31.99 & 4.00\\
		\hline
		Co-Ru swap (12.5\%) & 0.97, 0.96,  & 0.08, 0.05, & 2.99, 2.92, & $\mathrm{Co_{Ru}}$: 1.08,& 32.00 & 4.00\\
		& 0.98, 0.97, & 0.08, 0.18  & 3.02, 2.90,  & $\mathrm{Ru_{Co}}$: -0.16 &\\
		& 0.96, 1.00, & 0.09, 0.05 & 2.92, 3.00, & & & \\
		& 0.98 & 0.02 & 2.92, 3.00 & & & \\

		\hline
		Co-Ru swap (25.0\%) & 0.99, 1.00 & 0.04, 0.01, & 3.00, 2.90, & $\mathrm{Co_{Ru}}$: 1.09,1.09& 31.99 & 4.00\\
		& 1.00, 1.00 & 0.10, 0.10, & 2.97, 2.96,  & $\mathrm{Ru_{Co}}$: -0.10,-0.10 &\\
		&1.00, 0.99  & 0.04, 0.01 & 2.97, 2.96, &-0.03 & & \\
		& &  & 2.99, 3.00 & & & \\
		\hline
		
		Co-Ru swap (37.5\%) & 1.00, 1.02, & -0.01, 0.04,  & 2.98, 2.95,  & $\mathrm{Co_{Ru}}$: 1.07,1.07,1.08& 31.99 & 4.00\\
		& 1.02, 1.02 & 0.04, 0.05 & 2.97, 2.94, & $\mathrm{Ru_{Co}}$: -0.05, -0.05, -0.06 &\\
		& 1.07 & 0.00 & 2.99, 2.96,  & & \\
		& &  & 2.90, 2.95 & & & \\
		\hline
		Co-Ru swap (50\%) & 1.06, 1.06,  & -0.01, -0.04 & 2.96, 2.96, & $\mathrm{Co_{Ru}}$: 1.05,1.06,1.06,1.05& 32.00 & 4.00\\
		& 1.05, 1.05& -0.02, -0.01  & 2.96, 2.96,  & $\mathrm{Ru_{Co}}$: -0.01,-0.01,-0.02,-0.04 &\\
		&  &  & 2.97, 2.97, & & \\
		& &  &2.94, 2.94  & & & \\
		
		\hline \hline
	\end{tabular}
	\label{dis1}
\end{table*}
\textit{Hybridization of molecular levels in CoRuMnSi}: 
Atomic orbital projected density of states in the spin-resolved band structure helps to understand the hybridization of different atomic orbitals in the alloy. This helps to get a better insight into the properties of the alloy. For example, CoFeMnSi was reported to be spin gapless semiconductor while simply replacing Fe with Ru changes the nature from SGS to half metal. Figure \ref{fig:proj-orb-CoRuMnSi} shows the d-orbital characters of various atoms in CoRuMnSi along high symmetry K-directions of the spin-resolved band structure. The interaction of atomic orbitals in Heusler alloys was explained in previous literature. \cite{PhysRevB.66.174429,Graf-simplerules-Heusler-PSSC-review} For example, in Co$_2$MnSi alloy Co ions occupy tetrahedral sites while Mn occupies the octahedral site. Galanakis \textit{et. al.}\cite{PhysRevB.66.174429} considered tetrahedral site Co atomic $d$ orbitals to split as low lying $t_{2g}$ and higher-lying $e_g$ levels whereas Graf \textit{et. al.},\cite{Graf-simplerules-Heusler-PSSC-review} considered the initial levels in another way. The initial $t_{2g}$ and $e_g$ orbitals of each tetrahedral site Co was considered to hybridize and form covalent bonds, whereas the octahedral site Mn $d$ orbitals were reported to split into low lying $t_{2g}$ and higher-lying $e_g$ levels due to crystal field. These octahedral site $t_{2g}$ and $e_g$ states interact with other levels but retain their atomic character in the molecular picture (as these levels do not hybridize due to the ionic character). Similar schematics of molecular interaction picture can also be noticed in FeRhCrGe alloys.\cite{Venkateswara-FeRhCrGe-arxiv1902.01593v2} However, in CoRuMnSi alloy, we can notice the anomalies in the interaction/hybridization of atomic orbitals by carefully analyzing the projected atomic $d$ orbital character at $\Gamma$ point (see Fig. \ref{fig:proj-orb-CoRuMnSi}). If one focuses on the bands at $\mathrm{E_F}$ near the $\Gamma$-point in Fig.\ref{fig:proj-orb-CoRuMnSi} (a) and \ref{fig:proj-orb-CoRuMnSi}(c), by symmetry, theses should be the $e_g$ character of the hybridized orbitals arising out of Co and Ru-atoms which occupy tetrahedral sites. However, our simulation reveals that $e_g$-orbitals of Co instead interact with $e_g$-orbitals of octahedral site atoms (i.e., Mn) for spin-up band (see Fig.\ref{fig:proj-orb-CoRuMnSi}(a) and (e)). These anomalies leave the $e_g$ character at $E_F$ to be purely arising from the $e_g$ orbital character of Ru ions only, in the spin-up band structure. A similar nature of anomalies can also be noticed in the spin-down band structure, but instead of $e_g$ orbitals it occurs in $t_{2g}$ orbitals. Notice from Fig. \ref{fig:proj-orb-CoRuMnSi}(b) and (f) that low the $t_{2g}$ orbitals of Co hybridize with $t_{2g}$ orbitals of Mn in the spin-down band. Hence, we conclude that Co and Mn are forming some kind of unconventional bonding in CoRuMnSi leaving spin-up band structure purely dominated by Ru $e_g$ orbitals at the Fermi level, which could dominate in electrical and spin transport. Analyzing the projected atomic orbital character along high symmetry $K$-points of CoFeMnSi, one can conclude an identical nature of the interaction of atomic orbitals as observed in CoRuMnSi. Spin-up orbitals character of CoFeMnSi projected on to $d$ atomic orbitals of Fe are shown in Fig. \ref{fig:proj-3dFe-CoFeMnSi}. Hence, it is evident that the change of nature of $e_g$ orbitals at the Fermi level in the spin-up band structure is the underlying reason for the change of SGS behavior to localized half-metallic nature on replacing Fe with Ru. Notably, the difference lies in the nature of $e_g$ orbitals that are electron-like with light charge carriers in CoFeMnSi, unlike hole-like nature with heavy charge carriers in CoRuMnSi. Hence, we conclude that SGS nature observed in CoFeMnSi is due to the unique nature of $e_g$ orbitals of Fe. In contrast, the large density of states in the spin-up band structure of CoRuMnSi is due to the presence of well-localized flat hole-like $e_g$ orbitals at the Fermi level that are purely due to Ru states. So the unique localized half-metallic nature observed in CoRuMnSi is due to its unique band structure, and should not be constructed as due to other reasons such as disorder, impurities, etc.

\textit{Disorder}: The XRD analysis reveals that 4c and 4d fcc sites in CoRuMnSi are equally possible for Co and Ru atoms. To understand the electronic and magnetic properties of the actual experimental structure, we have simulated the electronic structure of the disordered configuration by considering a mixture of Co and Ru atoms at 4c and 4d sites. Swap analysis (equal probability) is one such way to investigate such disorder. To simulate this, a $2\times2\times2$ supercell involving 32-atoms is constructed, which contains 8-formula units. In such a cell, exchanging the position of 1 Co and 1 Ru gives a 12.5 \% swap disorder. Similarly, exchanging 2,3,4 Co and Ru atoms gives 25 \%, 37.5 \%, and 50 \% swap disorder, respectively. 50 \% swap (equal probability) corresponds to $\mathrm{L2_1}$-type structure. We have checked all possible configurations for replacement of Co by Ru and vice versa at each concentration, and results of the most stable configuration are presented here. The calculated spin-polarized DoS for disordered structures (with different percentages of swap disorder) are shown in Fig. \ref{dis}. These DoS indicate that half-metallicity in CoRuMnSi is robust against swapping. The total magnetization value almost remains the same as the ideal case (no defects). To get a better understanding, we have calculated the local moments at/near the individual atomic sites, as shown in Table \ref{dis1}. In each case, the magnetic moment of Ru swapped with Co atom changes from positive to negative sign (though by a very small moment) as compared to the ideal case, whereas the moment of Co swapped with Ru atoms slightly increases. With swap disorder, the local environment of the defect sites remains the same (Co and Ru atoms surrounded by 4 Mn and 4 Si atoms in both ordered and defect structures) with slightly different bond lengths (due to relaxation effect). Thus, the exchange interactions are not affected much, modifying the local moments (and thus the total moment) only by a very small amount. Co-Ru swap defect structures have nearly the same total moment as the ideal structure, 32 $\mu_B$/cell or 4 $\mu_B$/f.u.(see Table \ref {dis1}).

\section{Conclusion}
In conclusion, we have studied the effect of full replacement of Fe in equiatomic quaternary Heusler alloy CoFeMnSi with Ru by a combined experimental and theoretical investigation. Replacement of Fe with Ru improves the structural ordering from DO$_3$ to L2$_1$ as revealed by the room temperature XRD. The saturation magnetization is found to be 3.92 $\mu_B/f.u.$ at 3 K and is closely consistent with the Slater-Pauling rule. The Curie temperature is found to be 780 K. The resistivity vs. temperature behavior indirectly indicates the possibility of half-metallic nature (even at high temperatures), since one-magnon scattering ($\mathrm{T^2}$- dependence) is found to be absent. At lower temperatures, the resistivity is found to decrease with temperature followed by an upturn, which is attributed to the localization effect. The localization observed in the alloy arises purely from the intrinsic electronic band structure and more specifically from the spin-up band. 

\textit{Ab-initio} simulation confirms the half-metallic nature of CoRuMnSi with a clear band gap in the minority band. Spin-resolved band structure reveals that large density of states at the Fermi level in the spin-up band arises due to the presence of flat $e_g$ bands at $E_F$, which also produces heavy hole pockets at $\Gamma$. Spin-up band structure resembles that of a conventional localized metal, i.e., superposition of flat valence bands and sharp conduction bands. Hence, the localization effect seen in the resistivity measurements is purely attributed to the intrinsic spin up band structure of CoRuMnSi. A careful analysis of the projected atomic d-orbital character of spin-resolved band structure reveals the unusual hybridization of orbitals in CoRuMnSi, which is also identical to CoFeMnSi, unlike other Heusler alloys. The observed hybridization in these alloys predict the $e_g$ orbitals character, present at the Fermi level in the spin-up band structure, to be purely arising from either Fe or Ru.

To get a deeper insight into the disorder at tetrahedral site atoms as revealed from XRD, Co-Ru swap disordered structures have also been simulated. It is found that the swap disorder does not affect the electronic properties and half-metallicity. The local environment of the defect sites remains the same as that of the ordered structure with a small change in the bond lengths. This gives a minimal change in the exchange interactions and hence the total magnetic moment. Robust half-metallic nature, stable structure, and high $\mathrm{T_C}$ make this alloy quite promising to be used in spintronic applications.

\section*{AcknowledgmentS}
Y. V. acknowledges the financial help provided by IIT Bombay and SERB. D. R. would like to thank Council of Scientific and Industrial Research (CSIR), India for providing Senior Research Fellowship. Y.V. also thanks Dr. P. D. Babu, BARC for carrying out high-temperature magnetization measurements.

\bibliographystyle{apsrev4-1}
\bibliography{bib}

\begin{thebibliography}{45}%
\makeatletter
\providecommand \@ifxundefined [1]{%
 \@ifx{#1\undefined}
}%
\providecommand \@ifnum [1]{%
 \ifnum #1\expandafter \@firstoftwo
 \else \expandafter \@secondoftwo
 \fi
}%
\providecommand \@ifx [1]{%
 \ifx #1\expandafter \@firstoftwo
 \else \expandafter \@secondoftwo
 \fi
}%
\providecommand \natexlab [1]{#1}%
\providecommand \enquote  [1]{``#1''}%
\providecommand \bibnamefont  [1]{#1}%
\providecommand \bibfnamefont [1]{#1}%
\providecommand \citenamefont [1]{#1}%
\providecommand \href@noop [0]{\@secondoftwo}%
\providecommand \href [0]{\begingroup \@sanitize@url \@href}%
\providecommand \@href[1]{\@@startlink{#1}\@@href}%
\providecommand \@@href[1]{\endgroup#1\@@endlink}%
\providecommand \@sanitize@url [0]{\catcode `\\12\catcode `\$12\catcode
  `\&12\catcode `\#12\catcode `\^12\catcode `\_12\catcode `\%12\relax}%
\providecommand \@@startlink[1]{}%
\providecommand \@@endlink[0]{}%
\providecommand \url  [0]{\begingroup\@sanitize@url \@url }%
\providecommand \@url [1]{\endgroup\@href {#1}{\urlprefix }}%
\providecommand \urlprefix  [0]{URL }%
\providecommand \Eprint [0]{\href }%
\providecommand \doibase [0]{http://dx.doi.org/}%
\providecommand \selectlanguage [0]{\@gobble}%
\providecommand \bibinfo  [0]{\@secondoftwo}%
\providecommand \bibfield  [0]{\@secondoftwo}%
\providecommand \translation [1]{[#1]}%
\providecommand \BibitemOpen [0]{}%
\providecommand \bibitemStop [0]{}%
\providecommand \bibitemNoStop [0]{.\EOS\space}%
\providecommand \EOS [0]{\spacefactor3000\relax}%
\providecommand \BibitemShut  [1]{\csname bibitem#1\endcsname}%
\let\auto@bib@innerbib\@empty
\bibitem [{\citenamefont {Bainsla}\ \emph
  {et~al.}(2015{\natexlab{a}})\citenamefont {Bainsla}, \citenamefont {Mallick},
  \citenamefont {Raja}, \citenamefont {Nigam}, \citenamefont {Varaprasad},
  \citenamefont {Takahashi}, \citenamefont {Alam}, \citenamefont {Suresh},\
  and\ \citenamefont {Hono}}]{PhysRevB.91.104408}%
  \BibitemOpen
  \bibfield  {author} {\bibinfo {author} {\bibfnamefont {L.}~\bibnamefont
  {Bainsla}}, \bibinfo {author} {\bibfnamefont {A.~I.}\ \bibnamefont
  {Mallick}}, \bibinfo {author} {\bibfnamefont {M.~M.}\ \bibnamefont {Raja}},
  \bibinfo {author} {\bibfnamefont {A.~K.}\ \bibnamefont {Nigam}}, \bibinfo
  {author} {\bibfnamefont {B.~S. D. C.~S.}\ \bibnamefont {Varaprasad}},
  \bibinfo {author} {\bibfnamefont {Y.~K.}\ \bibnamefont {Takahashi}}, \bibinfo
  {author} {\bibfnamefont {A.}~\bibnamefont {Alam}}, \bibinfo {author}
  {\bibfnamefont {K.~G.}\ \bibnamefont {Suresh}}, \ and\ \bibinfo {author}
  {\bibfnamefont {K.}~\bibnamefont {Hono}},\ }\href {\doibase
  10.1103/PhysRevB.91.104408} {\bibfield  {journal} {\bibinfo  {journal} {Phys.
  Rev. B}\ }\textbf {\bibinfo {volume} {91}},\ \bibinfo {pages} {104408}
  (\bibinfo {year} {2015}{\natexlab{a}})}\BibitemShut {NoStop}%
\bibitem [{\citenamefont {Enamullah}\ \emph {et~al.}(2015)\citenamefont
  {Enamullah}, \citenamefont {Venkateswara}, \citenamefont {Gupta},
  \citenamefont {Varma}, \citenamefont {Singh}, \citenamefont {Suresh},\ and\
  \citenamefont {Alam}}]{PhysRevB.92.224413}%
  \BibitemOpen
  \bibfield  {author} {\bibinfo {author} {\bibnamefont {Enamullah}}, \bibinfo
  {author} {\bibfnamefont {Y.}~\bibnamefont {Venkateswara}}, \bibinfo {author}
  {\bibfnamefont {S.}~\bibnamefont {Gupta}}, \bibinfo {author} {\bibfnamefont
  {M.~R.}\ \bibnamefont {Varma}}, \bibinfo {author} {\bibfnamefont
  {P.}~\bibnamefont {Singh}}, \bibinfo {author} {\bibfnamefont {K.~G.}\
  \bibnamefont {Suresh}}, \ and\ \bibinfo {author} {\bibfnamefont
  {A.}~\bibnamefont {Alam}},\ }\href {\doibase 10.1103/PhysRevB.92.224413}
  {\bibfield  {journal} {\bibinfo  {journal} {Phys. Rev. B}\ }\textbf {\bibinfo
  {volume} {92}},\ \bibinfo {pages} {224413} (\bibinfo {year}
  {2015})}\BibitemShut {NoStop}%
\bibitem [{\citenamefont {Makinistian}\ \emph {et~al.}(2013)\citenamefont
  {Makinistian}, \citenamefont {Faiz}, \citenamefont {Panguluri}, \citenamefont
  {Balke}, \citenamefont {Wurmehl}, \citenamefont {Felser}, \citenamefont
  {Albanesi}, \citenamefont {Petukhov},\ and\ \citenamefont
  {Nadgorny}}]{PhysRevB.87.220402}%
  \BibitemOpen
  \bibfield  {author} {\bibinfo {author} {\bibfnamefont {L.}~\bibnamefont
  {Makinistian}}, \bibinfo {author} {\bibfnamefont {M.~M.}\ \bibnamefont
  {Faiz}}, \bibinfo {author} {\bibfnamefont {R.~P.}\ \bibnamefont {Panguluri}},
  \bibinfo {author} {\bibfnamefont {B.}~\bibnamefont {Balke}}, \bibinfo
  {author} {\bibfnamefont {S.}~\bibnamefont {Wurmehl}}, \bibinfo {author}
  {\bibfnamefont {C.}~\bibnamefont {Felser}}, \bibinfo {author} {\bibfnamefont
  {E.~A.}\ \bibnamefont {Albanesi}}, \bibinfo {author} {\bibfnamefont {A.~G.}\
  \bibnamefont {Petukhov}}, \ and\ \bibinfo {author} {\bibfnamefont
  {B.}~\bibnamefont {Nadgorny}},\ }\href {\doibase 10.1103/PhysRevB.87.220402}
  {\bibfield  {journal} {\bibinfo  {journal} {Phys. Rev. B}\ }\textbf {\bibinfo
  {volume} {87}},\ \bibinfo {pages} {220402} (\bibinfo {year}
  {2013})}\BibitemShut {NoStop}%
\bibitem [{\citenamefont {Bainsla}\ \emph
  {et~al.}(2015{\natexlab{b}})\citenamefont {Bainsla}, \citenamefont {Mallick},
  \citenamefont {Raja}, \citenamefont {Coelho}, \citenamefont {Nigam},
  \citenamefont {Johnson}, \citenamefont {Alam},\ and\ \citenamefont
  {Suresh}}]{PhysRevB.92.045201}%
  \BibitemOpen
  \bibfield  {author} {\bibinfo {author} {\bibfnamefont {L.}~\bibnamefont
  {Bainsla}}, \bibinfo {author} {\bibfnamefont {A.~I.}\ \bibnamefont
  {Mallick}}, \bibinfo {author} {\bibfnamefont {M.~M.}\ \bibnamefont {Raja}},
  \bibinfo {author} {\bibfnamefont {A.~A.}\ \bibnamefont {Coelho}}, \bibinfo
  {author} {\bibfnamefont {A.~K.}\ \bibnamefont {Nigam}}, \bibinfo {author}
  {\bibfnamefont {D.~D.}\ \bibnamefont {Johnson}}, \bibinfo {author}
  {\bibfnamefont {A.}~\bibnamefont {Alam}}, \ and\ \bibinfo {author}
  {\bibfnamefont {K.~G.}\ \bibnamefont {Suresh}},\ }\href {\doibase
  10.1103/PhysRevB.92.045201} {\bibfield  {journal} {\bibinfo  {journal} {Phys.
  Rev. B}\ }\textbf {\bibinfo {volume} {92}},\ \bibinfo {pages} {045201}
  (\bibinfo {year} {2015}{\natexlab{b}})}\BibitemShut {NoStop}%
\bibitem [{\citenamefont {Ouardi}\ \emph {et~al.}(2013)\citenamefont {Ouardi},
  \citenamefont {Fecher}, \citenamefont {Felser},\ and\ \citenamefont
  {K\"ubler}}]{PhysRevLett.110.100401}%
  \BibitemOpen
  \bibfield  {author} {\bibinfo {author} {\bibfnamefont {S.}~\bibnamefont
  {Ouardi}}, \bibinfo {author} {\bibfnamefont {G.~H.}\ \bibnamefont {Fecher}},
  \bibinfo {author} {\bibfnamefont {C.}~\bibnamefont {Felser}}, \ and\ \bibinfo
  {author} {\bibfnamefont {J.}~\bibnamefont {K\"ubler}},\ }\href {\doibase
  10.1103/PhysRevLett.110.100401} {\bibfield  {journal} {\bibinfo  {journal}
  {Phys. Rev. Lett.}\ }\textbf {\bibinfo {volume} {110}},\ \bibinfo {pages}
  {100401} (\bibinfo {year} {2013})}\BibitemShut {NoStop}%
\bibitem [{\citenamefont {Venkateswara}\ \emph {et~al.}(2018)\citenamefont
  {Venkateswara}, \citenamefont {Gupta}, \citenamefont {Samatham},
  \citenamefont {Varma}, \citenamefont {Enamullah}, \citenamefont {Suresh},\
  and\ \citenamefont {Alam}}]{PhysRevB.97.054407}%
  \BibitemOpen
  \bibfield  {author} {\bibinfo {author} {\bibfnamefont {Y.}~\bibnamefont
  {Venkateswara}}, \bibinfo {author} {\bibfnamefont {S.}~\bibnamefont {Gupta}},
  \bibinfo {author} {\bibfnamefont {S.~S.}\ \bibnamefont {Samatham}}, \bibinfo
  {author} {\bibfnamefont {M.~R.}\ \bibnamefont {Varma}}, \bibinfo {author}
  {\bibnamefont {Enamullah}}, \bibinfo {author} {\bibfnamefont {K.~G.}\
  \bibnamefont {Suresh}}, \ and\ \bibinfo {author} {\bibfnamefont
  {A.}~\bibnamefont {Alam}},\ }\href {\doibase 10.1103/PhysRevB.97.054407}
  {\bibfield  {journal} {\bibinfo  {journal} {Phys. Rev. B}\ }\textbf {\bibinfo
  {volume} {97}},\ \bibinfo {pages} {054407} (\bibinfo {year}
  {2018})}\BibitemShut {NoStop}%
\bibitem [{\citenamefont {Herran}\ \emph {et~al.}(2017)\citenamefont {Herran},
  \citenamefont {Dalal}, \citenamefont {Gray}, \citenamefont {Kharel},\ and\
  \citenamefont {Lukashev}}]{doi:10.1063/1.4998308}%
  \BibitemOpen
  \bibfield  {author} {\bibinfo {author} {\bibfnamefont {J.}~\bibnamefont
  {Herran}}, \bibinfo {author} {\bibfnamefont {R.}~\bibnamefont {Dalal}},
  \bibinfo {author} {\bibfnamefont {P.}~\bibnamefont {Gray}}, \bibinfo {author}
  {\bibfnamefont {P.}~\bibnamefont {Kharel}}, \ and\ \bibinfo {author}
  {\bibfnamefont {P.~V.}\ \bibnamefont {Lukashev}},\ }\href {\doibase
  10.1063/1.4998308} {\bibfield  {journal} {\bibinfo  {journal} {Journal of
  Applied Physics}\ }\textbf {\bibinfo {volume} {122}},\ \bibinfo {pages}
  {153904} (\bibinfo {year} {2017})}\BibitemShut {NoStop}%
\bibitem [{\citenamefont {Stinshoff}\ \emph {et~al.}(2017)\citenamefont
  {Stinshoff}, \citenamefont {Fecher}, \citenamefont {Chadov}, \citenamefont
  {Nayak}, \citenamefont {Balke}, \citenamefont {Ouardi}, \citenamefont
  {Nakamura},\ and\ \citenamefont {Felser}}]{doi:10.1063/1.5000351}%
  \BibitemOpen
  \bibfield  {author} {\bibinfo {author} {\bibfnamefont {R.}~\bibnamefont
  {Stinshoff}}, \bibinfo {author} {\bibfnamefont {G.~H.}\ \bibnamefont
  {Fecher}}, \bibinfo {author} {\bibfnamefont {S.}~\bibnamefont {Chadov}},
  \bibinfo {author} {\bibfnamefont {A.~K.}\ \bibnamefont {Nayak}}, \bibinfo
  {author} {\bibfnamefont {B.}~\bibnamefont {Balke}}, \bibinfo {author}
  {\bibfnamefont {S.}~\bibnamefont {Ouardi}}, \bibinfo {author} {\bibfnamefont
  {T.}~\bibnamefont {Nakamura}}, \ and\ \bibinfo {author} {\bibfnamefont
  {C.}~\bibnamefont {Felser}},\ }\href {\doibase 10.1063/1.5000351} {\bibfield
  {journal} {\bibinfo  {journal} {AIP Advances}\ }\textbf {\bibinfo {volume}
  {7}},\ \bibinfo {pages} {105009} (\bibinfo {year} {2017})}\BibitemShut
  {NoStop}%
\bibitem [{\citenamefont {Jamer}\ \emph {et~al.}(2017)\citenamefont {Jamer},
  \citenamefont {Wang}, \citenamefont {Stephen}, \citenamefont {McDonald},
  \citenamefont {Grutter}, \citenamefont {Sterbinsky}, \citenamefont {Arena},
  \citenamefont {Borchers}, \citenamefont {Kirby}, \citenamefont {Lewis},
  \citenamefont {Barbiellini}, \citenamefont {Bansil},\ and\ \citenamefont
  {Heiman}}]{PhysRevApplied.7.064036}%
  \BibitemOpen
  \bibfield  {author} {\bibinfo {author} {\bibfnamefont {M.~E.}\ \bibnamefont
  {Jamer}}, \bibinfo {author} {\bibfnamefont {Y.~J.}\ \bibnamefont {Wang}},
  \bibinfo {author} {\bibfnamefont {G.~M.}\ \bibnamefont {Stephen}}, \bibinfo
  {author} {\bibfnamefont {I.~J.}\ \bibnamefont {McDonald}}, \bibinfo {author}
  {\bibfnamefont {A.~J.}\ \bibnamefont {Grutter}}, \bibinfo {author}
  {\bibfnamefont {G.~E.}\ \bibnamefont {Sterbinsky}}, \bibinfo {author}
  {\bibfnamefont {D.~A.}\ \bibnamefont {Arena}}, \bibinfo {author}
  {\bibfnamefont {J.~A.}\ \bibnamefont {Borchers}}, \bibinfo {author}
  {\bibfnamefont {B.~J.}\ \bibnamefont {Kirby}}, \bibinfo {author}
  {\bibfnamefont {L.~H.}\ \bibnamefont {Lewis}}, \bibinfo {author}
  {\bibfnamefont {B.}~\bibnamefont {Barbiellini}}, \bibinfo {author}
  {\bibfnamefont {A.}~\bibnamefont {Bansil}}, \ and\ \bibinfo {author}
  {\bibfnamefont {D.}~\bibnamefont {Heiman}},\ }\href {\doibase
  10.1103/PhysRevApplied.7.064036} {\bibfield  {journal} {\bibinfo  {journal}
  {Phys. Rev. Applied}\ }\textbf {\bibinfo {volume} {7}},\ \bibinfo {pages}
  {064036} (\bibinfo {year} {2017})}\BibitemShut {NoStop}%
\bibitem [{\citenamefont {Wang}\ \emph {et~al.}(2018)\citenamefont {Wang},
  \citenamefont {Li}, \citenamefont {Cheng}, \citenamefont {Wang},\ and\
  \citenamefont {Chen}}]{doi:10.1063/1.5042604}%
  \BibitemOpen
  \bibfield  {author} {\bibinfo {author} {\bibfnamefont {X.}~\bibnamefont
  {Wang}}, \bibinfo {author} {\bibfnamefont {T.}~\bibnamefont {Li}}, \bibinfo
  {author} {\bibfnamefont {Z.}~\bibnamefont {Cheng}}, \bibinfo {author}
  {\bibfnamefont {X.-L.}\ \bibnamefont {Wang}}, \ and\ \bibinfo {author}
  {\bibfnamefont {H.}~\bibnamefont {Chen}},\ }\href {\doibase
  10.1063/1.5042604} {\bibfield  {journal} {\bibinfo  {journal} {Applied
  Physics Reviews}\ }\textbf {\bibinfo {volume} {5}},\ \bibinfo {pages}
  {041103} (\bibinfo {year} {2018})}\BibitemShut {NoStop}%
\bibitem [{\citenamefont {Singh}\ \emph {et~al.}(2018)\citenamefont {Singh},
  \citenamefont {Kashyap},\ and\ \citenamefont {Saini}}]{SINGH201815421}%
  \BibitemOpen
  \bibfield  {author} {\bibinfo {author} {\bibfnamefont {M.}~\bibnamefont
  {Singh}}, \bibinfo {author} {\bibfnamefont {M.~K.}\ \bibnamefont {Kashyap}},
  \ and\ \bibinfo {author} {\bibfnamefont {H.~S.}\ \bibnamefont {Saini}},\
  }\href {\doibase https://doi.org/10.1016/j.matpr.2018.05.027} {\bibfield
  {journal} {\bibinfo  {journal} {Materials Today: Proceedings}\ }\textbf
  {\bibinfo {volume} {5}},\ \bibinfo {pages} {15421 } (\bibinfo {year}
  {2018})},\ \bibinfo {note} {10th National Conference on Solid State Chemistry
  and Allied Areas, 1-3 July, 2017}\BibitemShut {NoStop}%
\bibitem [{\citenamefont {Picozzi}\ and\ \citenamefont
  {Freeman}(2007)}]{Picozzi_2007}%
  \BibitemOpen
  \bibfield  {author} {\bibinfo {author} {\bibfnamefont {S.}~\bibnamefont
  {Picozzi}}\ and\ \bibinfo {author} {\bibfnamefont {A.~J.}\ \bibnamefont
  {Freeman}},\ }\href {\doibase 10.1088/0953-8984/19/31/315215} {\bibfield
  {journal} {\bibinfo  {journal} {Journal of Physics: Condensed Matter}\
  }\textbf {\bibinfo {volume} {19}},\ \bibinfo {pages} {315215} (\bibinfo
  {year} {2007})}\BibitemShut {NoStop}%
\bibitem [{\citenamefont {Seema}\ and\ \citenamefont
  {Kumar}(2015)}]{doi:10.1063/1.4929252}%
  \BibitemOpen
  \bibfield  {author} {\bibinfo {author} {\bibfnamefont {K.}~\bibnamefont
  {Seema}}\ and\ \bibinfo {author} {\bibfnamefont {R.}~\bibnamefont {Kumar}},\
  }\href {\doibase 10.1063/1.4929252} {\bibfield  {journal} {\bibinfo
  {journal} {AIP Conference Proceedings}\ }\textbf {\bibinfo {volume} {1675}},\
  \bibinfo {pages} {030036} (\bibinfo {year} {2015})}\BibitemShut {NoStop}%
\bibitem [{\citenamefont {Bruski}\ \emph {et~al.}(2011)\citenamefont {Bruski},
  \citenamefont {Erwin}, \citenamefont {Ramsteiner}, \citenamefont {Brandt},
  \citenamefont {Friedland}, \citenamefont {Farshchi}, \citenamefont
  {Herfort},\ and\ \citenamefont {Riechert}}]{PhysRevB.83.140409}%
  \BibitemOpen
  \bibfield  {author} {\bibinfo {author} {\bibfnamefont {P.}~\bibnamefont
  {Bruski}}, \bibinfo {author} {\bibfnamefont {S.~C.}\ \bibnamefont {Erwin}},
  \bibinfo {author} {\bibfnamefont {M.}~\bibnamefont {Ramsteiner}}, \bibinfo
  {author} {\bibfnamefont {O.}~\bibnamefont {Brandt}}, \bibinfo {author}
  {\bibfnamefont {K.-J.}\ \bibnamefont {Friedland}}, \bibinfo {author}
  {\bibfnamefont {R.}~\bibnamefont {Farshchi}}, \bibinfo {author}
  {\bibfnamefont {J.}~\bibnamefont {Herfort}}, \ and\ \bibinfo {author}
  {\bibfnamefont {H.}~\bibnamefont {Riechert}},\ }\href {\doibase
  10.1103/PhysRevB.83.140409} {\bibfield  {journal} {\bibinfo  {journal} {Phys.
  Rev. B}\ }\textbf {\bibinfo {volume} {83}},\ \bibinfo {pages} {140409}
  (\bibinfo {year} {2011})}\BibitemShut {NoStop}%
\bibitem [{\citenamefont {Miura}\ \emph {et~al.}(2004)\citenamefont {Miura},
  \citenamefont {Nagao},\ and\ \citenamefont {Shirai}}]{PhysRevB.69.144413}%
  \BibitemOpen
  \bibfield  {author} {\bibinfo {author} {\bibfnamefont {Y.}~\bibnamefont
  {Miura}}, \bibinfo {author} {\bibfnamefont {K.}~\bibnamefont {Nagao}}, \ and\
  \bibinfo {author} {\bibfnamefont {M.}~\bibnamefont {Shirai}},\ }\href
  {\doibase 10.1103/PhysRevB.69.144413} {\bibfield  {journal} {\bibinfo
  {journal} {Phys. Rev. B}\ }\textbf {\bibinfo {volume} {69}},\ \bibinfo
  {pages} {144413} (\bibinfo {year} {2004})}\BibitemShut {NoStop}%
\bibitem [{\citenamefont {Graf}\ \emph {et~al.}(2011)\citenamefont {Graf},
  \citenamefont {Felser},\ and\ \citenamefont
  {Parkin}}]{Graf-simplerules-Heusler-PSSC-review}%
  \BibitemOpen
  \bibfield  {author} {\bibinfo {author} {\bibfnamefont {T.}~\bibnamefont
  {Graf}}, \bibinfo {author} {\bibfnamefont {C.}~\bibnamefont {Felser}}, \ and\
  \bibinfo {author} {\bibfnamefont {S.~S.}\ \bibnamefont {Parkin}},\ }\href
  {\doibase https://doi.org/10.1016/j.progsolidstchem.2011.02.001} {\bibfield
  {journal} {\bibinfo  {journal} {Progress in Solid State Chemistry}\ }\textbf
  {\bibinfo {volume} {39}},\ \bibinfo {pages} {1 } (\bibinfo {year}
  {2011})}\BibitemShut {NoStop}%
\bibitem [{\citenamefont {Rani}\ \emph {et~al.}(2017)\citenamefont {Rani},
  \citenamefont {Enamullah}, \citenamefont {Suresh}, \citenamefont {Yadav},
  \citenamefont {Jha}, \citenamefont {Bhattacharyya}, \citenamefont {Varma},\
  and\ \citenamefont {Alam}}]{PhysRevB.96.184404}%
  \BibitemOpen
  \bibfield  {author} {\bibinfo {author} {\bibfnamefont {D.}~\bibnamefont
  {Rani}}, \bibinfo {author} {\bibnamefont {Enamullah}}, \bibinfo {author}
  {\bibfnamefont {K.~G.}\ \bibnamefont {Suresh}}, \bibinfo {author}
  {\bibfnamefont {A.~K.}\ \bibnamefont {Yadav}}, \bibinfo {author}
  {\bibfnamefont {S.~N.}\ \bibnamefont {Jha}}, \bibinfo {author} {\bibfnamefont
  {D.}~\bibnamefont {Bhattacharyya}}, \bibinfo {author} {\bibfnamefont {M.~R.}\
  \bibnamefont {Varma}}, \ and\ \bibinfo {author} {\bibfnamefont
  {A.}~\bibnamefont {Alam}},\ }\href {\doibase 10.1103/PhysRevB.96.184404}
  {\bibfield  {journal} {\bibinfo  {journal} {Phys. Rev. B}\ }\textbf {\bibinfo
  {volume} {96}},\ \bibinfo {pages} {184404} (\bibinfo {year}
  {2017})}\BibitemShut {NoStop}%
\bibitem [{\citenamefont {Kharel}\ \emph {et~al.}(2017)\citenamefont {Kharel},
  \citenamefont {Herran}, \citenamefont {Lukashev}, \citenamefont {Jin},
  \citenamefont {Waybright}, \citenamefont {Gilbert}, \citenamefont {Staten},
  \citenamefont {Gray}, \citenamefont {Valloppilly}, \citenamefont {Huh},\ and\
  \citenamefont {Sellmyer}}]{doi:10.1063/1.4972797}%
  \BibitemOpen
  \bibfield  {author} {\bibinfo {author} {\bibfnamefont {P.}~\bibnamefont
  {Kharel}}, \bibinfo {author} {\bibfnamefont {J.}~\bibnamefont {Herran}},
  \bibinfo {author} {\bibfnamefont {P.}~\bibnamefont {Lukashev}}, \bibinfo
  {author} {\bibfnamefont {Y.}~\bibnamefont {Jin}}, \bibinfo {author}
  {\bibfnamefont {J.}~\bibnamefont {Waybright}}, \bibinfo {author}
  {\bibfnamefont {S.}~\bibnamefont {Gilbert}}, \bibinfo {author} {\bibfnamefont
  {B.}~\bibnamefont {Staten}}, \bibinfo {author} {\bibfnamefont
  {P.}~\bibnamefont {Gray}}, \bibinfo {author} {\bibfnamefont {S.}~\bibnamefont
  {Valloppilly}}, \bibinfo {author} {\bibfnamefont {Y.}~\bibnamefont {Huh}}, \
  and\ \bibinfo {author} {\bibfnamefont {D.~J.}\ \bibnamefont {Sellmyer}},\
  }\href {\doibase 10.1063/1.4972797} {\bibfield  {journal} {\bibinfo
  {journal} {AIP Advances}\ }\textbf {\bibinfo {volume} {7}},\ \bibinfo {pages}
  {056402} (\bibinfo {year} {2017})}\BibitemShut {NoStop}%
\bibitem [{\citenamefont {Choudhary}\ \emph {et~al.}(2016)\citenamefont
  {Choudhary}, \citenamefont {Kharel}, \citenamefont {Valloppilly},
  \citenamefont {Jin}, \citenamefont {O’Connell}, \citenamefont {Huh},
  \citenamefont {Gilbert}, \citenamefont {Kashyap}, \citenamefont {Sellmyer},\
  and\ \citenamefont {Skomski}}]{doi:10.1063/1.4943306}%
  \BibitemOpen
  \bibfield  {author} {\bibinfo {author} {\bibfnamefont {R.}~\bibnamefont
  {Choudhary}}, \bibinfo {author} {\bibfnamefont {P.}~\bibnamefont {Kharel}},
  \bibinfo {author} {\bibfnamefont {S.~R.}\ \bibnamefont {Valloppilly}},
  \bibinfo {author} {\bibfnamefont {Y.}~\bibnamefont {Jin}}, \bibinfo {author}
  {\bibfnamefont {A.}~\bibnamefont {O’Connell}}, \bibinfo {author}
  {\bibfnamefont {Y.}~\bibnamefont {Huh}}, \bibinfo {author} {\bibfnamefont
  {S.}~\bibnamefont {Gilbert}}, \bibinfo {author} {\bibfnamefont
  {A.}~\bibnamefont {Kashyap}}, \bibinfo {author} {\bibfnamefont {D.~J.}\
  \bibnamefont {Sellmyer}}, \ and\ \bibinfo {author} {\bibfnamefont
  {R.}~\bibnamefont {Skomski}},\ }\href {\doibase 10.1063/1.4943306} {\bibfield
   {journal} {\bibinfo  {journal} {AIP Advances}\ }\textbf {\bibinfo {volume}
  {6}},\ \bibinfo {pages} {056304} (\bibinfo {year} {2016})}\BibitemShut
  {NoStop}%
\bibitem [{\citenamefont {Alijani}\ \emph {et~al.}(2012)\citenamefont
  {Alijani}, \citenamefont {Winterlik}, \citenamefont {Fecher}, \citenamefont
  {Naghavi}, \citenamefont {Chadov}, \citenamefont {Gruhn},\ and\ \citenamefont
  {Felser}}]{Alijani_2012}%
  \BibitemOpen
  \bibfield  {author} {\bibinfo {author} {\bibfnamefont {V.}~\bibnamefont
  {Alijani}}, \bibinfo {author} {\bibfnamefont {J.}~\bibnamefont {Winterlik}},
  \bibinfo {author} {\bibfnamefont {G.~H.}\ \bibnamefont {Fecher}}, \bibinfo
  {author} {\bibfnamefont {S.~S.}\ \bibnamefont {Naghavi}}, \bibinfo {author}
  {\bibfnamefont {S.}~\bibnamefont {Chadov}}, \bibinfo {author} {\bibfnamefont
  {T.}~\bibnamefont {Gruhn}}, \ and\ \bibinfo {author} {\bibfnamefont
  {C.}~\bibnamefont {Felser}},\ }\href {\doibase 10.1088/0953-8984/24/4/046001}
  {\bibfield  {journal} {\bibinfo  {journal} {J. Phys.: Conden. Matt.}\
  }\textbf {\bibinfo {volume} {24}},\ \bibinfo {pages} {046001} (\bibinfo
  {year} {2012})}\BibitemShut {NoStop}%
\bibitem [{\citenamefont {Bainsla}\ \emph
  {et~al.}(2015{\natexlab{c}})\citenamefont {Bainsla}, \citenamefont {Raja},
  \citenamefont {Nigam},\ and\ \citenamefont {Suresh}}]{BAINSLA2015631}%
  \BibitemOpen
  \bibfield  {author} {\bibinfo {author} {\bibfnamefont {L.}~\bibnamefont
  {Bainsla}}, \bibinfo {author} {\bibfnamefont {M.~M.}\ \bibnamefont {Raja}},
  \bibinfo {author} {\bibfnamefont {A.}~\bibnamefont {Nigam}}, \ and\ \bibinfo
  {author} {\bibfnamefont {K.}~\bibnamefont {Suresh}},\ }\href {\doibase
  https://doi.org/10.1016/j.jallcom.2015.08.150} {\bibfield  {journal}
  {\bibinfo  {journal} {J. Alloy Compd.}\ }\textbf {\bibinfo {volume} {651}},\
  \bibinfo {pages} {631 } (\bibinfo {year} {2015}{\natexlab{c}})}\BibitemShut
  {NoStop}%
\bibitem [{\citenamefont {Klaer}\ \emph {et~al.}(2009)\citenamefont {Klaer},
  \citenamefont {Kallmayer}, \citenamefont {Elmers}, \citenamefont {Basit},
  \citenamefont {Thöne}, \citenamefont {Chadov},\ and\ \citenamefont
  {Felser}}]{Klaer_2009}%
  \BibitemOpen
  \bibfield  {author} {\bibinfo {author} {\bibfnamefont {P.}~\bibnamefont
  {Klaer}}, \bibinfo {author} {\bibfnamefont {M.}~\bibnamefont {Kallmayer}},
  \bibinfo {author} {\bibfnamefont {H.~J.}\ \bibnamefont {Elmers}}, \bibinfo
  {author} {\bibfnamefont {L.}~\bibnamefont {Basit}}, \bibinfo {author}
  {\bibfnamefont {J.}~\bibnamefont {Thöne}}, \bibinfo {author} {\bibfnamefont
  {S.}~\bibnamefont {Chadov}}, \ and\ \bibinfo {author} {\bibfnamefont
  {C.}~\bibnamefont {Felser}},\ }\href {\doibase 10.1088/0022-3727/42/8/084001}
  {\bibfield  {journal} {\bibinfo  {journal} {Journal of Physics D: Applied
  Physics}\ }\textbf {\bibinfo {volume} {42}},\ \bibinfo {pages} {084001}
  (\bibinfo {year} {2009})}\BibitemShut {NoStop}%
\bibitem [{\citenamefont {Kanomata}\ \emph {et~al.}(2006)\citenamefont
  {Kanomata}, \citenamefont {Kikuchi},\ and\ \citenamefont
  {Yamauchi}}]{KANOMATA20061}%
  \BibitemOpen
  \bibfield  {author} {\bibinfo {author} {\bibfnamefont {T.}~\bibnamefont
  {Kanomata}}, \bibinfo {author} {\bibfnamefont {M.}~\bibnamefont {Kikuchi}}, \
  and\ \bibinfo {author} {\bibfnamefont {H.}~\bibnamefont {Yamauchi}},\ }\href
  {\doibase https://doi.org/10.1016/j.jallcom.2005.06.079} {\bibfield
  {journal} {\bibinfo  {journal} {Journal of Alloys and Compounds}\ }\textbf
  {\bibinfo {volume} {414}},\ \bibinfo {pages} {1 } (\bibinfo {year}
  {2006})}\BibitemShut {NoStop}%
\bibitem [{\citenamefont {Kundu}\ \emph {et~al.}(2017)\citenamefont {Kundu},
  \citenamefont {Ghosh}, \citenamefont {Banerjee}, \citenamefont {Ghosh},\ and\
  \citenamefont {Sanyal}}]{Kundu-CoRuMnSi-theory-Scientificreports}%
  \BibitemOpen
  \bibfield  {author} {\bibinfo {author} {\bibfnamefont {A.}~\bibnamefont
  {Kundu}}, \bibinfo {author} {\bibfnamefont {S.}~\bibnamefont {Ghosh}},
  \bibinfo {author} {\bibfnamefont {R.}~\bibnamefont {Banerjee}}, \bibinfo
  {author} {\bibfnamefont {S.}~\bibnamefont {Ghosh}}, \ and\ \bibinfo {author}
  {\bibfnamefont {B.}~\bibnamefont {Sanyal}},\ }\href {\doibase
  10.1038/s41598-017-01782-5} {\bibfield  {journal} {\bibinfo  {journal}
  {Scientific Reports}\ }\textbf {\bibinfo {volume} {7}},\ \bibinfo {pages}
  {1803} (\bibinfo {year} {2017})}\BibitemShut {NoStop}%
\bibitem [{\citenamefont {Khalaf Al-zyadi}\ \emph {et~al.}(2018)\citenamefont
  {Khalaf Al-zyadi}, \citenamefont {Kadhim},\ and\ \citenamefont
  {Yao}}]{Khalaf-CoRuMnSi-theory-RSC}%
  \BibitemOpen
  \bibfield  {author} {\bibinfo {author} {\bibfnamefont {J.~M.}\ \bibnamefont
  {Khalaf Al-zyadi}}, \bibinfo {author} {\bibfnamefont {A.~A.}\ \bibnamefont
  {Kadhim}}, \ and\ \bibinfo {author} {\bibfnamefont {K.-L.}\ \bibnamefont
  {Yao}},\ }\href {\doibase 10.1039/C8RA02918K} {\bibfield  {journal} {\bibinfo
   {journal} {RSC Adv.}\ }\textbf {\bibinfo {volume} {8}},\ \bibinfo {pages}
  {25653} (\bibinfo {year} {2018})}\BibitemShut {NoStop}%
\bibitem [{RR()}]{RR}%
  \BibitemOpen
  \href {https://www.ill.eu/sites/fullprof/php/tutorials.html} {\enquote
  {\bibinfo {title} {https://www.ill.eu/sites/fullprof/php/tutorials.html},}\
  }\BibitemShut {NoStop}%
\bibitem [{\citenamefont
  {Rodríguez-Carvajal}(1993)}]{RODRIGUEZCARVAJAL199355}%
  \BibitemOpen
  \bibfield  {author} {\bibinfo {author} {\bibfnamefont {J.}~\bibnamefont
  {Rodríguez-Carvajal}},\ }\href {\doibase
  https://doi.org/10.1016/0921-4526(93)90108-I} {\bibfield  {journal} {\bibinfo
   {journal} {Physica B: Condensed Matter}\ }\textbf {\bibinfo {volume}
  {192}},\ \bibinfo {pages} {55 } (\bibinfo {year} {1993})}\BibitemShut
  {NoStop}%
\bibitem [{\citenamefont {Kresse}\ and\ \citenamefont
  {Furthm\"uller}(1996)}]{PhysRevB.54.11169}%
  \BibitemOpen
  \bibfield  {author} {\bibinfo {author} {\bibfnamefont {G.}~\bibnamefont
  {Kresse}}\ and\ \bibinfo {author} {\bibfnamefont {J.}~\bibnamefont
  {Furthm\"uller}},\ }\href {\doibase 10.1103/PhysRevB.54.11169} {\bibfield
  {journal} {\bibinfo  {journal} {Phys. Rev. B}\ }\textbf {\bibinfo {volume}
  {54}},\ \bibinfo {pages} {11169} (\bibinfo {year} {1996})}\BibitemShut
  {NoStop}%
\bibitem [{\citenamefont {Kresse}\ and\ \citenamefont
  {Joubert}(1999)}]{PhysRevB.59.1758}%
  \BibitemOpen
  \bibfield  {author} {\bibinfo {author} {\bibfnamefont {G.}~\bibnamefont
  {Kresse}}\ and\ \bibinfo {author} {\bibfnamefont {D.}~\bibnamefont
  {Joubert}},\ }\href {\doibase 10.1103/PhysRevB.59.1758} {\bibfield  {journal}
  {\bibinfo  {journal} {Phys. Rev. B}\ }\textbf {\bibinfo {volume} {59}},\
  \bibinfo {pages} {1758} (\bibinfo {year} {1999})}\BibitemShut {NoStop}%
\bibitem [{\citenamefont {Kokalji}(2003)}]{kokalji}%
  \BibitemOpen
  \bibfield  {author} {\bibinfo {author} {\bibfnamefont {A.}~\bibnamefont
  {Kokalji}},\ }\href {\doibase 10.1016/S0927-0256(03)00104-6} {\bibfield
  {journal} {\bibinfo  {journal} {Computational Materials Science}\ }\textbf
  {\bibinfo {volume} {28}},\ \bibinfo {pages} {155} (\bibinfo {year} {2003})},\
  \Eprint {http://arxiv.org/abs/Proceedings of the Symposium on Software
  Development for Process and Materials Design} {Proceedings of the Symposium
  on Software Development for Process and Materials Design} \BibitemShut
  {NoStop}%
\bibitem [{\citenamefont {Momma}\ and\ \citenamefont
  {Izumi}(2011)}]{Mommadb5098}%
  \BibitemOpen
  \bibfield  {author} {\bibinfo {author} {\bibfnamefont {K.}~\bibnamefont
  {Momma}}\ and\ \bibinfo {author} {\bibfnamefont {F.}~\bibnamefont {Izumi}},\
  }\href {\doibase 10.1107/S0021889811038970} {\bibfield  {journal} {\bibinfo
  {journal} {Journal of Applied Crystallography}\ }\textbf {\bibinfo {volume}
  {44}},\ \bibinfo {pages} {1272} (\bibinfo {year} {2011})}\BibitemShut
  {NoStop}%
\bibitem [{\citenamefont {Rani}\ \emph {et~al.}(2019)\citenamefont {Rani},
  \citenamefont {Enamullah}, \citenamefont {Bainsla}, \citenamefont {Suresh},\
  and\ \citenamefont {Alam}}]{PhysRevB.99.104429}%
  \BibitemOpen
  \bibfield  {author} {\bibinfo {author} {\bibfnamefont {D.}~\bibnamefont
  {Rani}}, \bibinfo {author} {\bibnamefont {Enamullah}}, \bibinfo {author}
  {\bibfnamefont {L.}~\bibnamefont {Bainsla}}, \bibinfo {author} {\bibfnamefont
  {K.~G.}\ \bibnamefont {Suresh}}, \ and\ \bibinfo {author} {\bibfnamefont
  {A.}~\bibnamefont {Alam}},\ }\href {\doibase 10.1103/PhysRevB.99.104429}
  {\bibfield  {journal} {\bibinfo  {journal} {Phys. Rev. B}\ }\textbf {\bibinfo
  {volume} {99}},\ \bibinfo {pages} {104429} (\bibinfo {year}
  {2019})}\BibitemShut {NoStop}%
\bibitem [{\citenamefont {Alijani}\ \emph {et~al.}(2011)\citenamefont
  {Alijani}, \citenamefont {Winterlik}, \citenamefont {Fecher}, \citenamefont
  {Naghavi},\ and\ \citenamefont {Felser}}]{PhysRevB.83.184428}%
  \BibitemOpen
  \bibfield  {author} {\bibinfo {author} {\bibfnamefont {V.}~\bibnamefont
  {Alijani}}, \bibinfo {author} {\bibfnamefont {J.}~\bibnamefont {Winterlik}},
  \bibinfo {author} {\bibfnamefont {G.~H.}\ \bibnamefont {Fecher}}, \bibinfo
  {author} {\bibfnamefont {S.~S.}\ \bibnamefont {Naghavi}}, \ and\ \bibinfo
  {author} {\bibfnamefont {C.}~\bibnamefont {Felser}},\ }\href {\doibase
  10.1103/PhysRevB.83.184428} {\bibfield  {journal} {\bibinfo  {journal} {Phys.
  Rev. B}\ }\textbf {\bibinfo {volume} {83}},\ \bibinfo {pages} {184428}
  (\bibinfo {year} {2011})}\BibitemShut {NoStop}%
\bibitem [{\citenamefont {Slater}(1936)}]{Slat1}%
  \BibitemOpen
  \bibfield  {author} {\bibinfo {author} {\bibfnamefont {J.~C.}\ \bibnamefont
  {Slater}},\ }\href {\doibase 10.1103/PhysRev.49.931} {\bibfield  {journal}
  {\bibinfo  {journal} {Phys. Rev.}\ }\textbf {\bibinfo {volume} {49}},\
  \bibinfo {pages} {931} (\bibinfo {year} {1936})}\BibitemShut {NoStop}%
\bibitem [{\citenamefont {Pauling}(1938)}]{Paul1}%
  \BibitemOpen
  \bibfield  {author} {\bibinfo {author} {\bibfnamefont {L.}~\bibnamefont
  {Pauling}},\ }\href {\doibase 10.1103/PhysRev.54.899} {\bibfield  {journal}
  {\bibinfo  {journal} {Phys. Rev.}\ }\textbf {\bibinfo {volume} {54}},\
  \bibinfo {pages} {899} (\bibinfo {year} {1938})}\BibitemShut {NoStop}%
\bibitem [{\citenamefont {Rani}\ \emph {et~al.}(2018)\citenamefont {Rani},
  \citenamefont {Kangsabanik}, \citenamefont {Suresh}, \citenamefont {Patra},
  \citenamefont {Bhattacharyya}, \citenamefont {Jha},\ and\ \citenamefont
  {Alam}}]{PhysRevApplied.10.054022}%
  \BibitemOpen
  \bibfield  {author} {\bibinfo {author} {\bibfnamefont {D.}~\bibnamefont
  {Rani}}, \bibinfo {author} {\bibfnamefont {J.}~\bibnamefont {Kangsabanik}},
  \bibinfo {author} {\bibfnamefont {K.~G.}\ \bibnamefont {Suresh}}, \bibinfo
  {author} {\bibfnamefont {N.}~\bibnamefont {Patra}}, \bibinfo {author}
  {\bibfnamefont {D.}~\bibnamefont {Bhattacharyya}}, \bibinfo {author}
  {\bibfnamefont {S.~N.}\ \bibnamefont {Jha}}, \ and\ \bibinfo {author}
  {\bibfnamefont {A.}~\bibnamefont {Alam}},\ }\href {\doibase
  10.1103/PhysRevApplied.10.054022} {\bibfield  {journal} {\bibinfo  {journal}
  {Phys. Rev. Applied}\ }\textbf {\bibinfo {volume} {10}},\ \bibinfo {pages}
  {054022} (\bibinfo {year} {2018})}\BibitemShut {NoStop}%
\bibitem [{\citenamefont {Aftab}\ \emph {et~al.}(2012)\citenamefont {Aftab},
  \citenamefont {Jaffari}, \citenamefont {Hasanain}, \citenamefont {Abbas},\
  and\ \citenamefont {Shah}}]{Aftab_2012}%
  \BibitemOpen
  \bibfield  {author} {\bibinfo {author} {\bibfnamefont {M.}~\bibnamefont
  {Aftab}}, \bibinfo {author} {\bibfnamefont {G.~H.}\ \bibnamefont {Jaffari}},
  \bibinfo {author} {\bibfnamefont {S.~K.}\ \bibnamefont {Hasanain}}, \bibinfo
  {author} {\bibfnamefont {T.~A.}\ \bibnamefont {Abbas}}, \ and\ \bibinfo
  {author} {\bibfnamefont {S.~I.}\ \bibnamefont {Shah}},\ }\href {\doibase
  10.1088/0022-3727/45/47/475001} {\bibfield  {journal} {\bibinfo  {journal}
  {Journal of Physics D: Applied Physics}\ }\textbf {\bibinfo {volume} {45}},\
  \bibinfo {pages} {475001} (\bibinfo {year} {2012})}\BibitemShut {NoStop}%
\bibitem [{\citenamefont {Srinivas}\ \emph {et~al.}(2013)\citenamefont
  {Srinivas}, \citenamefont {Prasanna~Kumari}, \citenamefont {Manivel~Raja},\
  and\ \citenamefont {Kamat}}]{doi:10.1063/1.4813519}%
  \BibitemOpen
  \bibfield  {author} {\bibinfo {author} {\bibfnamefont {K.}~\bibnamefont
  {Srinivas}}, \bibinfo {author} {\bibfnamefont {T.}~\bibnamefont
  {Prasanna~Kumari}}, \bibinfo {author} {\bibfnamefont {M.}~\bibnamefont
  {Manivel~Raja}}, \ and\ \bibinfo {author} {\bibfnamefont {S.~V.}\
  \bibnamefont {Kamat}},\ }\href {\doibase 10.1063/1.4813519} {\bibfield
  {journal} {\bibinfo  {journal} {Journal of Applied Physics}\ }\textbf
  {\bibinfo {volume} {114}},\ \bibinfo {pages} {033911} (\bibinfo {year}
  {2013})}\BibitemShut {NoStop}%
\bibitem [{\citenamefont {{Otto}}\ \emph {et~al.}(1989)\citenamefont {{Otto}},
  \citenamefont {{van Woerden}}, \citenamefont {{van der Valk}}, \citenamefont
  {{Wijngaard}}, \citenamefont {{van Bruggen}},\ and\ \citenamefont
  {{Haas}}}]{1989JPCM1.2351O}%
  \BibitemOpen
  \bibfield  {author} {\bibinfo {author} {\bibfnamefont {M.~J.}\ \bibnamefont
  {{Otto}}}, \bibinfo {author} {\bibfnamefont {R.~A.~M.}\ \bibnamefont {{van
  Woerden}}}, \bibinfo {author} {\bibfnamefont {P.~J.}\ \bibnamefont {{van der
  Valk}}}, \bibinfo {author} {\bibfnamefont {J.}~\bibnamefont {{Wijngaard}}},
  \bibinfo {author} {\bibfnamefont {C.~F.}\ \bibnamefont {{van Bruggen}}}, \
  and\ \bibinfo {author} {\bibfnamefont {C.}~\bibnamefont {{Haas}}},\ }\href
  {\doibase 10.1088/0953-8984/1/13/008} {\bibfield  {journal} {\bibinfo
  {journal} {J. Phys : Condens. Matter}\ }\textbf {\bibinfo {volume} {1}},\
  \bibinfo {pages} {2351} (\bibinfo {year} {1989})}\BibitemShut {NoStop}%
\bibitem [{\citenamefont {Kubo}(1972)}]{Kubo-1972}%
  \BibitemOpen
  \bibfield  {author} {\bibinfo {author} {\bibfnamefont {N.}~\bibnamefont
  {Kubo}, \bibfnamefont {Kenn;~Ohata}},\ }\href {\doibase 10.1143/JPSJ.33.21}
  {\bibfield  {journal} {\bibinfo  {journal} {J. Phys. Soc. Jpn.}\ }\textbf
  {\bibinfo {volume} {33}} (\bibinfo {year} {1972}),\
  10.1143/JPSJ.33.21}\BibitemShut {NoStop}%
\bibitem [{\citenamefont {Prestigiacomo}\ \emph {et~al.}(2014)\citenamefont
  {Prestigiacomo}, \citenamefont {Young}, \citenamefont {Adams},\ and\
  \citenamefont {Stadler}}]{doi:10.1063/1.4862966}%
  \BibitemOpen
  \bibfield  {author} {\bibinfo {author} {\bibfnamefont {J.~C.}\ \bibnamefont
  {Prestigiacomo}}, \bibinfo {author} {\bibfnamefont {D.~P.}\ \bibnamefont
  {Young}}, \bibinfo {author} {\bibfnamefont {P.~W.}\ \bibnamefont {Adams}}, \
  and\ \bibinfo {author} {\bibfnamefont {S.}~\bibnamefont {Stadler}},\ }\href
  {\doibase 10.1063/1.4862966} {\bibfield  {journal} {\bibinfo  {journal} {J.
  Appl. Phys.}\ }\textbf {\bibinfo {volume} {115}},\ \bibinfo {pages} {043712}
  (\bibinfo {year} {2014})}\BibitemShut {NoStop}%
\bibitem [{\citenamefont {Bainsla}\ \emph {et~al.}(2014)\citenamefont
  {Bainsla}, \citenamefont {Suresh}, \citenamefont {Nigam}, \citenamefont
  {Raja}, \citenamefont {Varaprasad}, \citenamefont {Takahashi},\ and\
  \citenamefont {Hono}}]{doi:10.1063/1.4902831}%
  \BibitemOpen
  \bibfield  {author} {\bibinfo {author} {\bibfnamefont {L.}~\bibnamefont
  {Bainsla}}, \bibinfo {author} {\bibfnamefont {K.~G.}\ \bibnamefont {Suresh}},
  \bibinfo {author} {\bibfnamefont {A.~K.}\ \bibnamefont {Nigam}}, \bibinfo
  {author} {\bibfnamefont {M.~M.}\ \bibnamefont {Raja}}, \bibinfo {author}
  {\bibfnamefont {B.~S. D. C.~S.}\ \bibnamefont {Varaprasad}}, \bibinfo
  {author} {\bibfnamefont {Y.~K.}\ \bibnamefont {Takahashi}}, \ and\ \bibinfo
  {author} {\bibfnamefont {K.}~\bibnamefont {Hono}},\ }\href {\doibase
  10.1063/1.4902831} {\bibfield  {journal} {\bibinfo  {journal} {J.
  Appl.Phys.}\ }\textbf {\bibinfo {volume} {116}},\ \bibinfo {pages} {203902}
  (\bibinfo {year} {2014})}\BibitemShut {NoStop}%
\bibitem [{\citenamefont {Gardelis}\ \emph {et~al.}(2004)\citenamefont
  {Gardelis}, \citenamefont {Androulakis}, \citenamefont {Migiakis},
  \citenamefont {Giapintzakis}, \citenamefont {Clowes}, \citenamefont
  {Bugoslavsky}, \citenamefont {Branford}, \citenamefont {Miyoshi},\ and\
  \citenamefont {Cohen}}]{doi:10.1063/1.1739293}%
  \BibitemOpen
  \bibfield  {author} {\bibinfo {author} {\bibfnamefont {S.}~\bibnamefont
  {Gardelis}}, \bibinfo {author} {\bibfnamefont {J.}~\bibnamefont
  {Androulakis}}, \bibinfo {author} {\bibfnamefont {P.}~\bibnamefont
  {Migiakis}}, \bibinfo {author} {\bibfnamefont {J.}~\bibnamefont
  {Giapintzakis}}, \bibinfo {author} {\bibfnamefont {S.~K.}\ \bibnamefont
  {Clowes}}, \bibinfo {author} {\bibfnamefont {Y.}~\bibnamefont {Bugoslavsky}},
  \bibinfo {author} {\bibfnamefont {W.~R.}\ \bibnamefont {Branford}}, \bibinfo
  {author} {\bibfnamefont {Y.}~\bibnamefont {Miyoshi}}, \ and\ \bibinfo
  {author} {\bibfnamefont {L.~F.}\ \bibnamefont {Cohen}},\ }\href {\doibase
  10.1063/1.1739293} {\bibfield  {journal} {\bibinfo  {journal} {J. Appl.
  Phys.}\ }\textbf {\bibinfo {volume} {95}},\ \bibinfo {pages} {8063} (\bibinfo
  {year} {2004})}\BibitemShut {NoStop}%
\bibitem [{\citenamefont {Venkateswara}\ \emph {et~al.}(2019)\citenamefont
  {Venkateswara}, \citenamefont {Samatham}, \citenamefont {Babu}, \citenamefont
  {Suresh},\ and\ \citenamefont
  {Alam}}]{Venkateswara-FeRhCrGe-arxiv1902.01593v2}%
  \BibitemOpen
  \bibfield  {author} {\bibinfo {author} {\bibfnamefont {Y.}~\bibnamefont
  {Venkateswara}}, \bibinfo {author} {\bibfnamefont {S.~S.}\ \bibnamefont
  {Samatham}}, \bibinfo {author} {\bibfnamefont {P.~D.}\ \bibnamefont {Babu}},
  \bibinfo {author} {\bibfnamefont {K.~G.}\ \bibnamefont {Suresh}}, \ and\
  \bibinfo {author} {\bibfnamefont {A.}~\bibnamefont {Alam}},\ }\href@noop {}
  {\enquote {\bibinfo {title} {Co-existence of spin semi-metallic and weyl
  semi-metallic behavior in ferhcrge},}\ } (\bibinfo {year} {2019}),\ \Eprint
  {http://arxiv.org/abs/arXiv:1902.01593} {arXiv:1902.01593} \BibitemShut
  {NoStop}%
\bibitem [{\citenamefont {Galanakis}\ \emph {et~al.}(2002)\citenamefont
  {Galanakis}, \citenamefont {Dederichs},\ and\ \citenamefont
  {Papanikolaou}}]{PhysRevB.66.174429}%
  \BibitemOpen
  \bibfield  {author} {\bibinfo {author} {\bibfnamefont {I.}~\bibnamefont
  {Galanakis}}, \bibinfo {author} {\bibfnamefont {P.~H.}\ \bibnamefont
  {Dederichs}}, \ and\ \bibinfo {author} {\bibfnamefont {N.}~\bibnamefont
  {Papanikolaou}},\ }\href {\doibase 10.1103/PhysRevB.66.174429} {\bibfield
  {journal} {\bibinfo  {journal} {Phys. Rev. B}\ }\textbf {\bibinfo {volume}
  {66}},\ \bibinfo {pages} {174429} (\bibinfo {year} {2002})}\BibitemShut
  {NoStop}%
\end{thebibliography}%

\end{document}